\documentclass[12pt,preprint]{aastex}
\usepackage{ulem}
\usepackage{longtable}
\usepackage{graphicx}

\shorttitle{The multi-color blackbody in GRB 081221}
\shortauthors{Hou et al.}
\begin{document}

\title{Multi-color blackbody emission in GRB 081221}
\author{Shu-Jin Hou\altaffilmark{1,2,3,4}, Bin-Bin Zhang\altaffilmark{5,6},
Yan-Zhi Meng\altaffilmark{2}, Xue-Feng Wu\altaffilmark{2}, En-Wei Liang%
\altaffilmark{3}, Hou-Jun L\"{u}\altaffilmark{3}, Tong Liu\altaffilmark{7},
Yun-Feng Liang\altaffilmark{2}, Lin Lin\altaffilmark{8},Rui-jing Lu%
\altaffilmark{3}, Jin-Shu Huang\altaffilmark{1,4}, and Bing Zhang%
\altaffilmark{9}}

\begin{abstract}
The radiation mechanism of the prompt emission of gamma-ray bursts (GRBs)
remains an open question. Although their spectra are usually well fitted
with the empirical Band function, which is widely believed to be fully
non-thermal and interpreted as an optically thin synchrotron emission,
accumulating evidence shows that a thermal component actually exists. In
this paper, a multi-color blackbody (mBB) model is proposed for the
time-integrated spectrum of GRB 081221 by assuming a power-law distribution
of the thermal luminosities with temperature, which manifests photospheric
emissions from  a different radius and/or angle. The effects of the minimum
temperature $kT_{\min }$, the maximum temperature $kT_{\max }$ and the power
law index $m$ of the luminosity distribution of an mBB are discussed.
The fitting to the time-integrated spectrum during the bright phase (from 20s to 30s since the trigger) of GRB 081221 by the mBB model
yields $kT_{\min }$ = $4.4\pm 0.3$ keV, $kT_{\max }$ = 57.0 $_{-1.4}^{+1.6}$
keV, and $m=-0.46_{-0.06}^{+0.05}$.
When the time bin is  small enough, the time-resolved spectra of GRB 081221 are well fitted with a series of single-temperature blackbodies.
Our results imply the prompt emission of GRB
081221 is dominated by the photosphere emission and its time-integrated
spectrum is a superposition of pure blackbody components at different times,
indicating that some empirical Band spectra may be interpreted as
mBB if the temperature is widely distributed.
\end{abstract}


\altaffiltext{1}{College of Physics and Electronic Engineering, Nanyang
Normal University, Nanyang, Henan, 473061, China}
\altaffiltext{2}{Purple
Mountain Observatory, Chinese Academy of Sciences, Nanjing 210008, China;
xfwu@pmo.ac.cn}
\altaffiltext{3}{Department of Physics and GXU-NAOC Center
for Astrophysics and Space Sciences, Guangxi University, Nanning 530004,
China; lew@gxu.edu.cn}
\altaffiltext{4}{Key Laboratory for the Structure and
Evolution of Celestial Objects, Chinese Academy of Sciences, Kunming 650011;
China}
\altaffiltext{5}{School of Astronomy and Space Science, Nanjing
University, Nanjing 210093, China; bbzhang@nju.edu.cn}
\altaffiltext{6}{Key
Laboratory of Modern Astronomy and Astrophysics (Nanjing University),
Ministry of Education, China}
\altaffiltext{7}{Department of Astronomy,
Xiamen University, Xiamen, Fujian 361005, China}
\altaffiltext{8}{Department
of Astronomy, Beijing Normal University, Beijing 100875, China}
\altaffiltext{9}{Department of Physics and Astronomy, University of Nevada,
Las Vegas, NV 89154; zhang@physics.unlv.edu}

\section{Introduction\label{sec:intro}}

The prompt emission is the earliest detected signal of gamma-ray bursts
(GRBs). However, its physical origin, closely related to the ejecta
composition, energy dissipation, particle acceleration, and especially the
radiation mechanism (e.g., Zhang 2011; Gao \& Zhang 2015; B\'{e}gu\'{e} \&
Pe'er 2015; Pe'er et al. 2015), is still a very controversial topic after
more than forty years.

Thermal emission in the relativistic fireball model is a natural explanation
for the prompt emission (Goodman et al. 1986), since the optical depth at
the base of the outflow is much larger than unity and it well explains the
observed small dispersion of the sub-MeV peak and high prompt emission
efficiency (M\'{e}sz\'{a}ros \& Rees 2000, Rees \& M\'{e}sz\'{a}ros 2005).
But in Compton GRO/BATSE era, the observed GRB spectra are well
fitted by an empirical smoothly-jointed broken power law, which is
the so-called ``Band" function (Band et al. 1993). Compared to a blackbody
function, this spectrum has a much wider peak, and it is much softer at low
energies. For a long time, such a spectral shape is widely believed to be
produced by optically thin synchrotron processes. Band function is then
used to phenomenologically capture this shape. However, previous studies
using both Compton/BASTE and Fermi/GBM data have shown that some observed low-energy
spectra are found to be harder than that predicted by the synchrotron
model in a substantial fraction of GRBs (Preece et al. 1998; Kaneko et al.
2006; Gruber et al.2014) and the spectral width of some GRBs is found to be
very narrow (Axelsson \& Borgonovo 2015; Yu et al. 2015). These facts
suggest that a quasi-thermal component should exist, which has been
observationally confirmed lately. The early examples include tens of BATSE
GRBs (Ryde 2005, Ryde \& Pe'er 2009). Thanks to Fermi's unprecedented
spectral energy coverage (GBM: 8 keV to 40 MeV, Meegan et al. 2009; LAT: 20
MeV to $>$300 GeV, Atwood et al. 2009), we now have several such
examples among the Fermi detected GRBs, such as GRB 090902B (Abdo et al.
2009; Ryde et al. 2010; Zhang et al. 2011), GRB 100724B (Guiriec et al.
2011), GRB 110721A (Axelsson et al. 2012), GRB 100507 (Ghirlanda et al.
2013), GRB 120323A (Guiriec et al. 2013) and GRB 101219B (Larsson et al. 2015).
The thermal emission is also claimed to exist in some early X-ray flares of GRBs (Peng et al. 2014).

GRB 081221 was jointly detected by Fermi/GBM (Wilson-Hodge 2008), the Burst Alert
Telescope (BAT) on board Swift (Hoversten et al. 2008; Cummings et al.
2008), and Konus/Wind (Golenetskii et al. 2008). The X-ray telescope on
board Swift found the early-time X-ray afterglow (Stroh \& Hoversten 2008).
The faint optical and radio afterglows were also detected (Kuin \& Hoversten
2008; Chandra \& Frail 2008). Multi-wavelength observations of this burst
provide the opportunity to explore the physical origin of the GRB.
Especially, the spectral properties of its prompt emission are similar to
those of GRB 090902B (Pe'er \& Ryde 2011; Ryde et al. 2010; Mizuta et al.
2011; Toma et al. 2011), showing a prominent thermal component. In this
work, we perform both the time-integrated and time-resolved spectral
analysis of the prompt emission of GRB 081221, to achieve the characteristic
of the thermal component.

In \S 2, we give a short description of our model. The time-integrated and
time-resolved spectral analysis are presented in \S 3. Conclusions are drawn
in \S 4 with some discussion.

\section{Multi-color blackbody spectra}
In the standard fireball model of GRBs, the photons produced in the
outflow are thermalized, and they can escape near the photospheric radius $%
r_{\mathrm{ph}}$, where the optical depth is equal to 1. In a relativistic,
spherically symmetric wind, the photospheric radius is found to be a
function of the angle to the line of sight (Abramowicz et al. 1991; Pe'er
2008;Meng et al. 2018 ), $r_{\mathrm{ph}}=$ $r_{\mathrm{ph}}(\theta)$. This means that the
observed spectrum is a superposition of a series of blackbodies of different
temperature, arising from different angles to the line of sight. Furthermore,
considering that photons have a finite probability of being scattered at any
point in space where electrons exist, Pe'er (2008) introduced the
probability density distributions for the last scattering photon positions$,$
$P(r,\theta)$. Then, the observed spectrum is contributed by photons from
the entire emitting volume, so it is certainly a multi-color blackbody (mBB).
Several works have tried to explain the Band function as
originated from the mBB by taking into account more
realistic outflow (Pe'er \& Ryde 2011; Beloborodov et al. 2011; Lundman et
al. 2013; Ito et al. 2013; Deng \& Zhang 2014).

The photon spectrum of a cosmological GRB with a single temperature $%
kT$, with luminosity $L_{\mathrm{BB}}$ defined in the host galaxy frame and
the luminosity distance $D_{\mathrm{L}}$ can be given as
\begin{equation}
N(E)=\frac{15}{4\pi^5}\frac{L_{\mathrm{BB}}}{D_{\mathrm{L}}^2(kT)^{4}}\frac{%
E^{2}}{{e^{E/kT}}-1}.
\end{equation}
Technically, the formula for the photon spectrum of a blackbody usually used
in the spectral fitting is
\begin{equation}
N(E)=\frac{8.0525K}{(e^{E/kT}-1)}\left(\frac{kT}{\mathrm{keV}}%
\right)^{-4}\left(\frac{E}{\mathrm{keV}}\right)^{2},
\end{equation}
where $N(E)$ is in units of ph cm$^{-2}$ s$^{-1}$ keV$^{-1}$, and $kT$ and $%
E $ are measured in the observer's frame. $K=L_{39}/D^2_{L,\mathrm{10 kpc}}$
is defined by the blackbody luminosity $L$ in units of $10^{39}$ erg s$^{-1}$
in the GRB host galaxy frame and the luminosity distance $D_{L}$ in units of
10 kpc.

For mBB, different from the \textbf{probability distribution} of
the mBB temperature in Ryde et al. (2010), we assume that
the dependence\textbf{\ }of the luminosity distribution on the temperature
has a power-law form,
\begin{equation}
\frac{dL}{dT}=\frac{m+1}{\left[ \left( \frac{T_{\max }}{T_{\min }}\right)
^{m+1}-1\right] }\frac{L_{\mathrm{mBB}}}{T_{\min }}\left( \frac{T}{T_{\min }}%
\right) ^{m},
\end{equation}%
where $m$ is the power-law index of the distribution, and the temperature
ranges from the minimum $T_{\min }$ to the maximum $T_{\max }$, and$L_{%
\mathrm{mBB}}=\int_{T_{\min }}^{T_{\max }}dL$ is the total luminosity of the
mBB. Compared with the previous treatment, Eq. (3) is more
directly related to the physics of the GRB fireball, since it may imply the
dependence of the luminosity (rather than temperature) in different radius
or angle. Combining the above equations, the photon spectrum of the
mBB can be formulated as
\begin{equation}
N(E)=\frac{8.0525(m+1)K}{\left[ \left( \frac{T_{\max }}{T_{\min }}\right)
^{m+1}-1\right] }\left( \frac{kT_{\min }}{\mathrm{keV}}\right) ^{-2}I(E),
\end{equation}%
where
\begin{equation}
I(E)=\left( \frac{E}{kT_{\min }}\right) ^{m-1}\int_{\frac{E}{kT_{\max }}}^{%
\frac{E}{kT_{\min }}}\frac{x^{2-m}}{e^{x}-1}dx,
\end{equation}%
and $x=E/kT$. Again, $N(E)$ is in units of ph cm$^{-2}$ s$^{-1}$ keV$^{-1}$.

\begin{figure}[ptb]
\label{Fig_1} \centering\includegraphics[angle=0,scale=1.0]{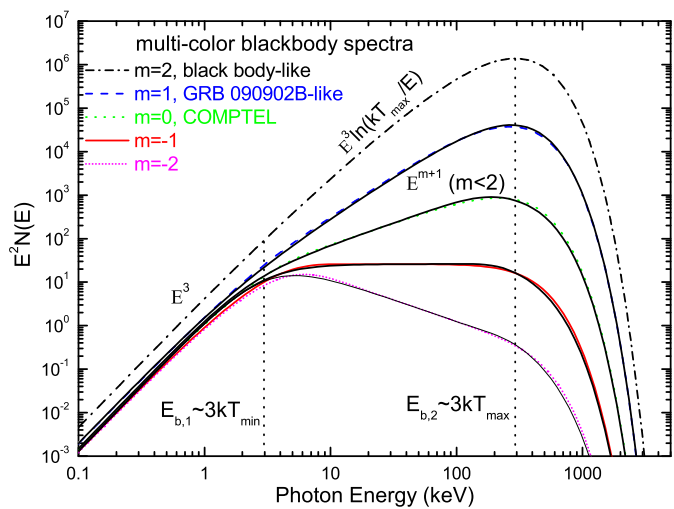}
\caption{Examples of the mBB spectra for different $m$. $kT_{\min}=1$ keV
and $kT_{\max}=100$ keV are assumed. Black solid lines are plotted with
analytic approximations Eq. (6), while other lines are exact numerical
calculations with Eq. (4). }
\end{figure}

%%%%%%%%%%%%%%%%%%%%%%%%%%%%%%%%%%Table 1%%%%%%%%%%%%%%%%%%%%%%%%%%%%%%%
\begin{table*}[tbp]
\caption{The fitting results of the time-resolved spectra of GRB 081221 with
the BB model and the mBB model }
\label{tab:specfitting1}
\begin{center}
\setlength{\tabcolsep}{2pt}
\begin{tiny}
\par
\begin{tabular}{ccccccccccc}
\hline
Time (s) & A$_{1}$ & $kT$ (keV) & PGSTAT/dof & BIC & A$_{2}$ & $kT_{\min }$
(keV) & $kT_{\max }$ (keV) & $m$ & PGSTAT/dof & BIC \\ \hline
(0.00,2.00) & $0.69_{-0.03}^{+0.03}$ & $30.72_{-1.48}^{+1.64}$ & 388.85/364
& 400.65 & $2.03_{-0.21}^{+2.16}$ & $3.74_{-0.36}^{+5.13}$ & $%
79.32_{-3.58}^{+26.08}$ & $0.25_{-0.54}^{+0.05}$ & 336.88/362 & 360.49 \\
(2.00,4.00) & $0.65_{-0.02}^{+0.02}$ & $16.44_{-0.56}^{+0.64}$ & 398.35/364
& 410.15 & $3.56_{-0.82}^{+0.66}$ & $5.64_{-1.61}^{+1.78}$ & $%
54.88_{-7.88}^{+17.04}$ & $-0.65_{-0.43}^{+0.39}$ & 308.02/362 & 331.63 \\
(4.00,6.00) & $0.43_{-0.03}^{+0.02}$ & $11.61_{-0.56}^{+0.64}$ & 365.33/364
& 377.12 & $3.70_{-0.08}^{+0.40}$ & $5.84_{-0.24}^{+1.14}$ & $%
46.49_{-13.07}^{+346.38}$ & $-1.31_{-0.18}^{+0.18}$ & 302.36/362 & 325.98 \\
(6.00,8.00) & $0.24_{-0.03}^{+0.03}$ & $9.65_{-0.63}^{+0.73}$ & 344.54/364 &
356.34 & $3.47_{-0.28}^{+0.31}$ & $5.49_{-0.68}^{+0.86}$ & $%
364.53_{-142.56}^{+493.58}$ & $-1.54_{-0.29}^{+0.17}$ & 313.94/362 & 337.55
\\
(8.00,10.00) & $0.29_{-0.03}^{+0.03}$ & $10.21_{-0.59}^{+0.64}$ & 299.41/364
& 311.20 & $3.01_{-0.34}^{+0.34}$ & $4.42_{-0.85}^{+0.85}$ & $%
36.11_{-16.54}^{+338.91}$ & $-1.16_{-0.23}^{+0.23}$ & 267.90/362 & 291.51 \\
(10.00,12.00) & $0.14_{-0.03}^{+0.03}$ & $7.16_{-0.46}^{+0.53}$ & 332.47/364
& 344.27 & $3.03_{-0.49}^{+0.14}$ & $4.36_{-0.96}^{+0.35}$ & $%
106.22_{-99.35}^{+345.39}$ & $-1.81_{-0.11}^{+0.29}$ & 317.17/362 & 340.78
\\
(12.00,14.00) & $0.21_{-0.03}^{+0.03}$ & $9.80_{-0.65}^{+0.84}$ & 328.09/364
& 339.89 & $2.91_{-0.53}^{+0.47}$ & $4.33_{-0.96}^{+1.08}$ & $%
157.80_{-118.79}^{+331.54}$ & $-1.19_{-0.24}^{+0.15}$ & 284.23/362 & 307.84
\\
(14.00,16.00) & $0.50_{-0.02}^{+0.02}$ & $10.21_{-0.42}^{+0.44}$ & 350.09/364
& 361.89 & $2.07_{-0.36}^{+0.36}$ & $2.62_{-0.76}^{+0.76}$ & $%
33.91_{-7.64}^{+378.53}$ & $-0.73_{-0.21}^{+0.21}$ & 267.42/362 & 291.04 \\
(16.00,17.00) & $0.44_{-0.03}^{+0.03}$ & $9.02_{-0.66}^{+0.72}$ & 370.33/364
& 382.13 & $2.83_{-0.23}^{+0.58}$ & $3.58_{-0.36}^{+1.17}$ & $%
64.27_{-34.99}^{+366.77}$ & $-1.29_{-0.30}^{+0.09}$ & 317.60/362 & 341.21 \\
(17.00,18.00) & $0.88_{-0.02}^{+0.02}$ & $15.35_{-0.53}^{+0.54}$ & 478.32/364
& 490.12 & $1.39_{-0.02}^{+1.70}$ & $1.87_{-0.02}^{+2.14}$ & $%
52.40_{-2.74}^{+11.31}$ & $-0.33_{-0.32}^{+0.01}$ & 317.05/362 & 340.66 \\
(18.00,18.50) & $1.16_{-0.02}^{+0.02}$ & $19.99_{-0.55}^{+0.66}$ & 329.25/364
& 341.05 & $1.17_{-0.39}^{+1.22}$ & $1.98_{-0.24}^{+2.28}$ & $%
44.59_{-1.20}^{+12.58}$ & $0.42_{-0.37}^{+0.37}$ & 261.68/362 & 285.29 \\
(18.50,19.00) & $1.26_{-0.02}^{+0.02}$ & $20.63_{-0.56}^{+0.58}$ & 370.22/364
& 382.02 & $3.61_{-0.02}^{+1.03}$ & $5.01_{-0.04}^{+2.52}$ & $%
55.19_{-2.83}^{+8.85}$ & $-0.06_{-0.45}^{+0.06}$ & 260.35/362 & 283.96 \\
(19.00,19.50) & $1.28_{-0.02}^{+0.01}$ & $21.29_{-0.60}^{+0.60}$ & 361.55/364
& 373.35 & $3.15_{-0.75}^{+0.88}$ & $4.33_{-1.16}^{+2.27}$ & $%
50.38_{-1.56}^{+14.58}$ & $0.30_{-0.45}^{+0.38}$ & 262.67/362 & 286.28 \\
(19.50,20.00) & $1.24_{-0.02}^{+0.02}$ & $19.93_{-0.50}^{+0.52}$ & 314.17/364
& 325.97 & $2.97_{-0.80}^{+0.92}$ & $4.12_{-1.19}^{+2.42}$ & $%
38.11_{-7.12}^{+8.36}$ & $0.61_{-0.67}^{+0.38}$ & 245.89/362 & 269.50 \\
(20.00,20.50) & $1.27_{-0.02}^{+0.02}$ & $20.09_{-0.51}^{+0.57}$ & 397.53/364
& 409.32 & $3.36_{-0.67}^{+1.03}$ & $4.46_{-1.06}^{+2.29}$ & $%
54.78_{-4.99}^{+7.30}$ & $-0.04_{-0.40}^{+0.18}$ & 269.52/362 & 293.13 \\
(20.50,21.00) & $1.35_{-0.01}^{+0.01}$ & $21.47_{-0.57}^{+0.54}$ & 473.93/364
& 485.73 & $2.97_{-0.11}^{+0.62}$ & $3.80_{-0.18}^{+1.41}$ & $%
55.92_{-0.64}^{+12.07}$ & $0.14_{-0.26}^{+0.26}$ & 314.18/362 & 337.79 \\
(21.00,21.50) & $1.43_{-0.01}^{+0.01}$ & $22.38_{-0.46}^{+0.51}$ & 405.52/364
& 417.32 & $2.46_{-0.21}^{+0.80}$ & $3.18_{-0.39}^{+1.55}$ & $%
48.25_{-5.50}^{+5.50}$ & $0.48_{-0.26}^{+0.26}$ & 254.51/362 & 278.12 \\
(21.50,22.00) & $1.42_{-0.01}^{+0.01}$ & $21.35_{-0.41}^{+0.48}$ & 384.52/364
& 396.32 & $2.30_{-0.18}^{+0.91}$ & $3.05_{-0.28}^{+1.93}$ & $%
40.79_{-1.35}^{+9.23}$ & $0.70_{-0.32}^{+0.26}$ & 284.85/362 & 308.46 \\
(22.00,22.50) & $1.32_{-0.01}^{+0.01}$ & $19.44_{-0.45}^{+0.48}$ & 421.93/364
& 433.73 & $2.75_{-0.27}^{+1.83}$ & $3.40_{-0.35}^{+3.62}$ & $%
48.13_{-2.12}^{+11.45}$ & $0.14_{-0.70}^{+0.06}$ & 268.20/362 & 291.81 \\
(22.50,23.00) & $1.22_{-0.02}^{+0.01}$ & $18.53_{-0.48}^{+0.57}$ & 348.99/364
& 360.79 & $1.60_{-0.71}^{+1.03}$ & $2.13_{-0.58}^{+1.58}$ & $%
46.69_{-1.62}^{+9.11}$ & $0.17_{-0.50}^{+0.02}$ & 234.22/362 & 257.83 \\
(23.00,23.50) & $1.18_{-0.02}^{+0.02}$ & $16.54_{-0.49}^{+0.53}$ & 389.35/364
& 401.15 & $4.35_{-0.45}^{+0.31}$ & $6.29_{-1.10}^{+0.87}$ & $%
63.59_{-6.73}^{+8.37}$ & $-0.82_{-0.21}^{+0.21}$ & 218.00/362 & 241.61 \\
(23.50,24.00) & $1.26_{-0.01}^{+0.01}$ & $18.46_{-0.46}^{+0.54}$ & 412.20/364
& 424.00 & $3.06_{-0.16}^{+1.14}$ & $3.79_{-0.26}^{+2.23}$ & $%
53.81_{-2.11}^{+10.00}$ & $-0.12_{-0.41}^{+0.03}$ & 256.61/362 & 280.22 \\
(24.00,24.50) & $1.36_{-0.01}^{+0.01}$ & $20.43_{-0.50}^{+0.49}$ & 414.72/364
& 426.52 & $3.47_{-1.08}^{+0.86}$ & $4.51_{-1.58}^{+1.86}$ & $%
55.93_{-5.05}^{+6.89}$ & $-0.05_{-0.33}^{+0.23}$ & 238.71/362 & 262.32 \\
(24.50,25.00) & $1.28_{-0.02}^{+0.01}$ & $17.85_{-0.42}^{+0.45}$ & 435.96/364
& 447.76 & $4.05_{-0.65}^{+0.45}$ & $5.53_{-1.30}^{+1.12}$ & $%
55.18_{-4.96}^{+6.54}$ & $-0.47_{-0.24}^{+0.26}$ & 262.46/362 & 286.07 \\
(25.00,25.50) & $1.16_{-0.02}^{+0.02}$ & $15.16_{-0.40}^{+0.47}$ & 434.11/364
& 445.91 & $2.30_{-0.63}^{+1.25}$ & $2.51_{-0.62}^{+1.84}$ & $%
47.81_{-3.07}^{+6.95}$ & $-0.35_{-0.30}^{+0.09}$ & 268.31/362 & 291.92 \\
(25.50,26.00) & $1.16_{-0.02}^{+0.01}$ & $14.18_{-0.41}^{+0.39}$ & 360.71/364
& 372.51 & $1.80_{-0.23}^{+0.80}$ & $2.09_{-0.26}^{+1.17}$ & $%
36.18_{-1.69}^{+6.04}$ & $-0.07_{-0.54}^{+0.02}$ & 241.03/362 & 264.64 \\
(26.00,26.50) & $1.15_{-0.02}^{+0.02}$ & $13.20_{-0.36}^{+0.40}$ & 390.14/364
& 401.94 & $4.47_{-0.19}^{+0.45}$ & $6.27_{-0.49}^{+1.31}$ & $%
51.84_{-3.05}^{+190.77}$ & $-1.19_{-0.44}^{+0.11}$ & 254.73/362 & 278.34 \\
(26.50,27.00) & $1.15_{-0.01}^{+0.01}$ & $13.18_{-0.36}^{+0.33}$ & 417.67/364
& 429.47 & $1.83_{-0.02}^{+1.60}$ & $2.02_{-0.04}^{+1.98}$ & $%
35.13_{-1.29}^{+5.61}$ & $-0.20_{-0.43}^{+0.02}$ & 278.27/362 & 301.88 \\
(27.00,27.50) & $1.16_{-0.02}^{+0.01}$ & $12.86_{-0.35}^{+0.40}$ & 406.79/364
& 418.59 & $4.38_{-0.33}^{+0.35}$ & $5.93_{-0.77}^{+0.99}$ & $%
55.82_{-6.50}^{+17.44}$ & $-1.21_{-0.26}^{+0.22}$ & 250.26/362 & 273.88 \\
(27.50,28.00) & $1.13_{-0.02}^{+0.02}$ & $13.33_{-0.43}^{+0.46}$ & 452.96/364
& 464.75 & $3.58_{-0.66}^{+0.41}$ & $4.24_{-1.07}^{+0.84}$ & $%
55.57_{-6.24}^{+8.64}$ & $-0.85_{-0.18}^{+0.22}$ & 252.09/362 & 275.70 \\
(28.00,28.50) & $1.09_{-0.02}^{+0.02}$ & $12.98_{-0.41}^{+0.45}$ & 417.06/364
& 428.86 & $3.63_{-0.38}^{+0.51}$ & $4.39_{-0.69}^{+1.09}$ & $%
53.55_{-4.45}^{+12.81}$ & $-0.87_{-0.27}^{+0.14}$ & 253.16/362 & 276.77 \\
(28.50,29.00) & $1.08_{-0.02}^{+0.02}$ & $13.19_{-0.48}^{+0.49}$ & 396.99/364
& 408.79 & $3.38_{-0.38}^{+0.59}$ & $3.96_{-0.59}^{+1.14}$ & $%
50.39_{-4.12}^{+10.44}$ & $-0.77_{-0.27}^{+0.13}$ & 235.18/362 & 258.79 \\
(29.00,29.50) & $1.03_{-0.02}^{+0.02}$ & $13.20_{-0.52}^{+0.55}$ & 398.43/364
& 410.24 & $3.44_{-0.45}^{+0.50}$ & $4.13_{-0.75}^{+1.04}$ & $%
62.45_{-6.83}^{+15.85}$ & $-0.86_{-0.21}^{+0.14}$ & 232.75/362 & 256.36 \\
(29.50,30.00) & $1.01_{-0.02}^{+0.02}$ & $12.45_{-0.46}^{+0.53}$ & 362.23/364
& 374.04 & $3.61_{-0.49}^{+0.51}$ & $4.47_{-0.86}^{+1.08}$ & $%
46.57_{-5.27}^{+10.44}$ & $-0.91_{-0.32}^{+0.19}$ & 245.11/362 & 268.72 \\
(30.00,30.50) & $0.97_{-0.02}^{+0.02}$ & $12.58_{-0.48}^{+0.49}$ & 335.44/364
& 347.24 & $3.56_{-0.50}^{+0.72}$ & $4.44_{-0.87}^{+1.64}$ & $%
52.72_{-8.14}^{+46.95}$ & $-0.94_{-0.43}^{+0.24}$ & 233.09/362 & 256.70 \\
(30.50,31.00) & $0.91_{-0.02}^{+0.02}$ & $11.69_{-0.54}^{+0.58}$ & 376.28/364
& 388.08 & $3.61_{-0.15}^{+0.67}$ & $4.55_{-0.30}^{+1.58}$ & $%
56.60_{-3.49}^{+405.12}$ & $-1.07_{-0.44}^{+0.06}$ & 276.00/362 & 299.61 \\
(31.00,31.50) & $0.86_{-0.02}^{+0.02}$ & $11.57_{-0.51}^{+0.55}$ & 283.93/364
& 295.73 & $3.94_{-0.01}^{+0.60}$ & $5.39_{-0.08}^{+1.63}$ & $%
43.76_{-15.08}^{+369.45}$ & $-1.22_{-0.24}^{+0.24}$ & 228.62/362 & 252.23 \\
(31.50,33.60) & $0.74_{-0.01}^{+0.01}$ & $11.30_{-0.34}^{+0.33}$ & 382.33/364
& 394.13 & $3.04_{-0.45}^{+1.06}$ & $3.86_{-0.65}^{+2.19}$ & $%
35.49_{-1.99}^{+20.33}$ & $-0.77_{-0.75}^{+0.18}$ & 271.64/362 & 295.25 \\
(33.60,35.20) & $0.54_{-0.02}^{+0.02}$ & $9.78_{-0.42}^{+0.41}$ & 383.78/364
& 395.58 & $3.35_{-0.23}^{+0.23}$ & $4.53_{-0.59}^{+0.59}$ & $%
38.59_{-11.85}^{+378.53}$ & $-1.29_{-0.17}^{+0.17}$ & 312.56/362 & 336.17 \\
(35.20,40.00) & $0.27_{-0.02}^{+0.02}$ & $8.99_{-0.31}^{+0.34}$ & 340.80/364
& 352.60 & $3.64_{-0.26}^{+0.05}$ & $5.76_{-0.68}^{+0.15}$ & $%
849.34_{-317.78}^{+150.64}$ & $-1.92_{-0.03}^{+0.21}$ & 297.39/362 & 321.00
\\ \hline
\end{tabular}%
\end{tiny}
\end{center}
\end{table*}
%%%%%%%%%%%%%%%%%%%%%%%%%%%table 1%%%%%%%%%%%%%%%%%%%%%%%%%%%%%%%%%%%%%%%

%%%%%%%%%%%%%%%%%%%%%%%%%%%table 2%%%%%%%%%%%%%%%%%%%%%%%%%%%%%%%%%%%%%%%

\clearpage
\begin{table*}[tbp]
\caption{The fitting results of the time-integrated spectrum of GRB 081221
from 20 s to 30 s using mBB and Band function model}
%\label{tab:specfitting3}
\begin{center}
%\begin{tiny}%
\par
\begin{tabular}{ccccc}
\hline
&  & mBB model &  &  \\ \hline
$kT_{\min }$ (keV) & $kT_{\max }$ (keV) & $m$ & PGSTAT/dof & BIC \\ \hline
$4.4\pm 0.3$ & $57.0$ $_{-1.4}^{+1.6}$ & $-0.46_{-0.06}^{+0.05}$ & 484.2/362
& 507.8 \\ \hline
&  & Band function model &  &  \\ \hline
$\alpha $ & $\beta $ & $E_{p}$ & PGSTAT/dof & BIC \\ \hline
$-0.60\pm 0.03$ & $-3.3_{-0.2}^{+0.1}$ & $90.1_{-1.2}^{+1.6}$ & 468.1/362 &
491.7 \\ \hline
\end{tabular}%
\end{center}
\end{table*}
%%%%%%%%%%%%%%%%%%%%%%%%%%%table 2%%%%%%%%%%%%%%%%%%%%%%%%%%%%%%%%%%%%%%%

%%%%%%%%%%%%%%%%%%%%%%%%%%%table 3%%%%%%%%%%%%%%%%%%%%%%%%%%%%%%%%%%%%%%%
\clearpage
\begin{deluxetable}{cccccc}
\tabletypesize{\scriptsize}
\tablecaption{The comparison of the BIC for the time-resolved spectral fitting of
GRB 081221 with the BB, the mBB and the 2BBPL models}
\tablenum{3}
\tablewidth{0pt}
\tabletypesize{\tiny}
\tablehead{
\colhead{Tine}&
\colhead{BIC$_{BB}$}&
\colhead{BIC$_{mBB}$}&
\colhead{BIC$_{2BBPL}$}&
\colhead{BIC$_{mBB}$ - BIC$_{BB}$}&
\colhead{BIC$_{2BBPL}$ - BIC$_{BB}$}}
\startdata
0.00$\sim$0.67&256.05&265.93&276.29&9.88&20.24 \\
0.67$\sim$0.77&163.06&174.96&186.09&11.90&23.03\\
0.77$\sim$0.87&146.14&158.49&169.80&12.35&23.66\\
0.87$\sim$0.97&132.97&144.91&156.45&11.94&23.48\\
0.97$\sim$1.07&154.86&166.57&178.07&11.71&23.21\\
1.07$\sim$1.17&159.68&171.59&183.08&11.91&23.40\\
1.17$\sim$1.27&166.20&179.41&189.74&13.21&23.54\\
1.27$\sim$1.37&153.06&164.81&176.52&11.75&23.46\\
1.37$\sim$1.47&160.48&172.48&183.82&11.99&23.33\\
1.47$\sim$1.57&166.02&176.93&189.38&10.91&23.36\\
1.57$\sim$1.67&155.17&168.85&178.64&13.68&23.47\\
1.67$\sim$1.77&142.25&154.01&165.73&11.76&23.48\\
1.77$\sim$1.87&142.53&154.27&165.90&11.74&23.37\\
1.87$\sim$1.97&148.46&162.00&171.95&13.55&23.50\\
1.97$\sim$2.07&131.41&141.83&154.16&10.42&22.75\\
2.07$\sim$2.17&127.25&138.93&150.73&11.69&23.48\\
2.17$\sim$2.27&134.52&144.83&156.27&10.31&21.75\\
2.27$\sim$2.37&144.38&156.05&166.83&11.67&22.45\\
2.37$\sim$2.47&152.27&163.91&175.74&11.65&23.48\\
2.47$\sim$2.57&151.96&163.37&173.70&11.40&21.73\\
2.57$\sim$2.67&138.23&150.16&161.57&11.93&23.34\\
2.67$\sim$2.77&139.66&150.58&162.34&10.92&22.68\\
2.77$\sim$2.87&150.93&162.21&174.41&11.28&23.48\\
2.87$\sim$2.97&166.39&177.64&188.73&11.25&22.34\\
2.97$\sim$3.07&151.20&162.62&174.47&11.41&23.26\\
3.07$\sim$3.17&148.46&160.01&171.94&11.56&23.48\\
3.17$\sim$3.27&147.78&159.31&171.26&11.53&23.49\\
3.27$\sim$3.37&136.96&148.69&160.46&11.73&23.50\\
3.37$\sim$3.47&136.30&147.95&159.81&11.64&23.51\\
3.47$\sim$3.57&164.98&176.69&188.46&11.71&23.49\\
3.57$\sim$3.67&131.49&143.23&154.99&11.74&23.50\\
3.67$\sim$3.77&138.22&149.92&161.71&11.69&23.49\\
3.77$\sim$3.87&160.07&171.02&183.32&10.95&23.25\\
3.87$\sim$3.97&161.91&173.64&185.39&11.73&23.48\\
3.97$\sim$4.07&151.72&163.42&175.09&11.70&23.37\\
4.07$\sim$4.67&275.38&271.56&285.75&-3.82&10.37\\
4.67$\sim$5.33&251.75&253.43&263.59&1.68&11.84\\
5.33$\sim$6.00&282.56&280.57&292.08&-1.99&9.52\\
6.00$\sim$6.67&255.33&253.63&264.14&-1.70&8.81\\
6.67$\sim$7.33&252.66&263.44&275.24&10.78&22.58\\
7.33$\sim$8.00&255.57&264.88&276.96&9.31&21.39\\
8.00$\sim$8.67&259.47&265.78&276.88&6.31&17.41\\
8.67$\sim$9.33&253.04&254.60&267.35&1.56&14.31\\
9.33$\sim$10.00&242.98&246.50&258.32&3.52&15.34\\
10.00$\sim$10.67&243.74&250.10&261.68&6.36&17.94\\
10.67$\sim$11.34&247.07&256.10&267.65&9.03&20.58\\
11.34$\sim$12.00&253.76&264.03&275.12&10.27&21.36\\
12.00$\sim$12.67&235.10&243.46&254.82&8.36&19.72\\
12.67$\sim$15.33&266.91&277.31&288.49&10.40&21.58\\
15.33$\sim$15.67&208.43&211.10&221.19&2.67&12.76\\
15.67$\sim$16.00&213.91&218.08&231.27&4.17&17.36\\
16.00$\sim$16.10&119.25&130.61&142.52&11.36&23.27\\
16.10$\sim$16.20&138.01&149.75&161.48&11.74&23.47\\
16.20$\sim$16.30&140.26&151.77&163.77&11.51&23.51\\
16.30$\sim$16.40&159.18&170.90&182.68&11.72&23.50\\
16.40$\sim$16.50&140.37&151.37&163.48&11.00&23.12\\
16.50$\sim$16.60&139.87&151.60&163.38&11.73&23.51\\
16.60$\sim$16.70&139.70&150.85&163.20&11.14&23.49\\
16.70$\sim$16.80&136.73&148.04&160.21&11.31&23.49\\
16.80$\sim$16.90&144.63&156.05&168.11&11.43&23.49\\
16.90$\sim$17.00&152.34&163.17&174.81&10.83&22.46\\
17.00$\sim$17.10&160.83&171.36&183.09&10.53&22.25\\
17.10$\sim$17.20&128.09&139.85&149.42&11.76&21.33\\
17.20$\sim$17.30&158.65&165.57&177.83&6.92&19.18\\
17.30$\sim$17.40&144.31&152.10&164.05&7.79&19.74\\
17.40$\sim$17.50&158.98&167.83&176.23&8.85&17.25\\
17.50$\sim$17.60&145.35&150.76&163.09&5.41&17.75\\
17.60$\sim$17.70&172.41&177.77&189.15&5.36&16.74\\
17.70$\sim$17.80&176.06&176.37&188.14&0.31&12.08\\
17.80$\sim$17.90&189.64&198.03&210.22&8.39&20.58\\
17.90$\sim$18.00&191.48&199.89&211.45&8.40&19.97\\
18.00$\sim$18.10&184.69&189.73&198.57&5.04&13.88\\
18.10$\sim$18.20&187.04&194.63&201.79&7.59&14.75\\
18.20$\sim$18.30&156.25&165.17&176.39&8.92&20.13\\
18.30$\sim$18.40&203.47&210.40&224.38&6.94&20.92\\
18.40$\sim$18.50&175.11&174.00&183.14&-1.11&8.03\\
18.50$\sim$18.60&209.72&211.57&223.23&1.86&13.51\\
18.60$\sim$18.70&151.38&158.14&168.22&6.76&16.84\\
18.70$\sim$18.80&238.87&235.80&245.64&-3.07&6.76\\
18.80$\sim$18.90&202.84&211.46&220.62&8.62&17.78\\
18.90$\sim$19.00&190.99&180.60&192.22&-10.39&1.24\\
19.00$\sim$19.10&170.73&175.92&186.47&5.19&15.74\\
19.10$\sim$19.20&205.92&208.74&220.56&2.82&14.64\\
19.20$\sim$19.30&220.96&213.03&223.20&-7.93&2.24\\
19.30$\sim$19.40&184.45&188.62&201.18&4.17&16.73\\
19.40$\sim$19.50&172.91&179.04&188.86&6.14&15.95\\
19.50$\sim$19.60&188.42&184.51&191.64&-3.91&3.22\\
19.60$\sim$19.70&196.12&206.00&219.07&9.88&22.95\\
19.70$\sim$19.80&154.11&156.86&166.20&2.75&12.09\\
19.80$\sim$19.90&171.93&181.54&194.16&9.61&22.23\\
19.90$\sim$20.00&182.33&188.80&202.42&6.47&20.09\\
20.00$\sim$20.10&178.06&177.95&186.95&-0.11&8.89\\
20.10$\sim$20.20&208.05&203.65&209.37&-4.39&1.32\\
20.20$\sim$20.30&218.22&214.45&224.90&-3.78&6.68\\
20.30$\sim$20.40&233.46&223.39&234.59&-10.07&1.13\\
20.40$\sim$20.50&207.90&212.39&222.89&4.48&14.98\\
20.50$\sim$20.60&222.75&199.74&208.77&-23.01&-13.97\\
20.60$\sim$20.70&204.40&196.08&209.56&-8.32&5.16\\
20.70$\sim$20.80&226.36&219.33&232.11&-7.02&5.76\\
20.80$\sim$20.90&203.05&209.95&221.64&6.90&18.59\\
20.90$\sim$21.00&220.08&207.14&214.90&-12.94&-5.18\\
21.00$\sim$21.10&207.37&207.83&214.21&0.46&6.84\\
21.10$\sim$21.20&191.65&177.20&189.19&-14.45&-2.46\\
21.20$\sim$21.30&187.38&181.57&190.85&-5.81&3.47\\
21.30$\sim$21.40&203.11&191.11&204.61&-12.00&1.50\\
21.40$\sim$21.50&198.27&189.27&200.72&-9.00&2.45\\
21.50$\sim$21.60&188.64&177.99&187.01&-10.66&-1.63\\
21.60$\sim$21.70&182.86&186.65&198.93&3.79&16.07\\
21.70$\sim$21.80&183.93&188.65&200.60&4.72&16.67\\
21.80$\sim$21.90&208.08&204.34&216.54&-3.74&8.46\\
21.90$\sim$22.00&197.73&202.44&206.14&4.71&8.41\\
22.00$\sim$22.10&217.01&214.35&220.98&-2.65&3.97\\
22.10$\sim$22.20&212.97&198.48&207.70&-14.49&-5.27\\
22.20$\sim$22.30&181.44&161.99&169.98&-19.45&-11.46\\
22.30$\sim$22.40&174.37&175.94&186.64&1.57&12.27\\
22.40$\sim$22.50&193.43&191.49&202.23&-1.94&8.79\\
22.50$\sim$22.60&174.69&177.55&184.81&2.86&10.12\\
22.60$\sim$22.70&154.70&159.05&171.59&4.35&16.89\\
22.70$\sim$22.80&170.25&178.47&190.53&8.22&20.29\\
22.80$\sim$22.90&194.25&181.45&192.00&-12.81&-2.25\\
22.90$\sim$23.00&174.42&169.57&182.26&-4.85&7.84\\
23.00$\sim$23.10&194.12&187.35&197.52&-6.77&3.40\\
23.10$\sim$23.20&193.26&189.46&200.54&-3.79&7.29\\
23.20$\sim$23.30&203.06&194.53&204.17&-8.53&1.11\\
23.30$\sim$23.40&180.35&172.15&183.64&-8.21&3.28\\
23.40$\sim$23.50&170.61&160.44&170.03&-10.16&-0.58\\
23.50$\sim$23.60&202.59&205.74&210.93&3.15&8.34\\
23.60$\sim$23.70&237.94&233.49&247.12&-4.45&9.18\\
23.70$\sim$23.80&190.02&180.35&190.95&-9.68&0.92\\
23.80$\sim$23.90&184.72&186.09&199.21&1.37&14.49\\
23.90$\sim$24.00&196.14&177.51&188.94&-18.64&-7.20\\
24.00$\sim$24.10&199.37&195.83&207.68&-3.54&8.32\\
24.10$\sim$24.20&190.29&193.12&202.97&2.83&12.68\\
24.20$\sim$24.30&226.08&202.10&215.47&-23.98&-10.60\\
24.30$\sim$24.40&216.09&199.08&212.84&-17.01&-3.26\\
24.40$\sim$24.50&196.75&181.18&189.75&-15.57&-7.00\\
24.50$\sim$24.60&216.32&205.19&213.01&-11.13&-3.31\\
24.60$\sim$24.70&213.81&195.80&210.60&-18.01&-3.21\\
24.70$\sim$24.80&194.98&185.41&195.29&-9.57&0.31\\
24.80$\sim$24.90&147.48&142.97&153.20&-4.51&5.72\\
24.90$\sim$25.00&183.59&175.05&187.22&-8.53&3.64\\
25.00$\sim$25.10&231.76&215.88&226.04&-15.88&-5.72\\
25.10$\sim$25.20&189.87&179.70&190.21&-10.16&0.34\\
25.20$\sim$25.30&172.13&171.41&183.89&-0.71&11.76\\
25.30$\sim$25.40&169.01&168.62&181.44&-0.40&12.43\\
25.40$\sim$25.50&179.88&173.96&184.37&-5.92&4.49\\
25.50$\sim$25.60&191.94&197.37&210.59&5.43&18.65\\
25.60$\sim$25.70&175.34&165.55&177.74&-9.79&2.41\\
25.70$\sim$25.80&183.28&185.36&195.75&2.08&12.47\\
25.80$\sim$25.90&213.34&217.37&228.02&4.04&14.69\\
25.90$\sim$26.00&163.26&164.65&174.48&1.39&11.22\\
26.00$\sim$26.10&151.56&154.81&167.56&3.25&16.00\\
26.10$\sim$26.20&197.51&187.90&198.73&-9.61&1.22\\
26.20$\sim$26.30&197.78&194.71&205.45&-3.08&7.67\\
26.30$\sim$26.40&214.50&205.95&215.38&-8.55&0.88\\
26.40$\sim$26.50&164.50&172.61&184.13&8.11&19.63\\
26.50$\sim$26.60&189.09&191.87&203.49&2.78&14.40\\
26.60$\sim$26.70&188.40&183.74&196.51&-4.65&8.11\\
26.70$\sim$26.80&158.44&159.42&173.14&0.97&14.69\\
26.80$\sim$26.90&184.99&187.09&199.69&2.10&14.70\\
26.90$\sim$27.00&190.92&187.11&200.57&-3.81&9.65\\
27.00$\sim$27.10&170.08&174.78&188.70&4.70&18.62\\
27.10$\sim$27.20&162.90&166.33&179.19&3.43&16.29\\
27.20$\sim$27.30&190.46&186.73&197.17&-3.72&6.72\\
27.30$\sim$27.40&183.42&177.64&189.68&-5.79&6.26\\
27.40$\sim$27.50&194.24&180.25&192.88&-13.99&-1.35\\
27.50$\sim$27.60&203.85&196.77&205.84&-7.08&1.98\\
27.60$\sim$27.70&181.29&162.55&171.49&-18.74&-9.80\\
27.70$\sim$27.80&200.48&186.88&197.85&-13.60&-2.63\\
27.80$\sim$27.90&180.43&167.74&176.31&-12.70&-4.12\\
27.90$\sim$28.00&174.39&168.61&179.39&-5.78&5.00\\
28.00$\sim$28.10&197.94&194.10&205.03&-3.84&7.09\\
28.10$\sim$28.20&198.42&186.08&202.35&-12.35&3.93\\
28.20$\sim$28.30&188.50&185.48&198.93&-3.02&10.43\\
28.30$\sim$28.40&194.75&199.06&210.73&4.31&15.98\\
28.40$\sim$28.50&195.05&187.07&199.93&-7.99&4.87\\
28.50$\sim$28.60&162.56&158.44&167.76&-4.12&5.20\\
28.60$\sim$28.70&194.32&184.54&196.62&-9.77&2.31\\
28.70$\sim$28.80&166.64&158.21&172.29&-8.43&5.66\\
28.80$\sim$28.90&159.29&160.38&172.09&1.09&12.80\\
28.90$\sim$29.00&189.64&188.83&201.01&-0.82&11.36\\
29.00$\sim$29.10&204.35&201.13&212.66&-3.21&8.31\\
29.10$\sim$29.20&201.35&201.25&213.75&-0.11&12.40\\
29.20$\sim$29.30&163.01&163.79&175.65&0.78&12.64\\
29.30$\sim$29.40&190.09&182.91&191.79&-7.18&1.70\\
29.40$\sim$29.50&193.32&187.47&201.17&-5.84&7.86\\
29.50$\sim$29.60&162.99&168.64&180.78&5.65&17.78\\
29.60$\sim$29.70&161.95&166.78&181.59&4.83&19.64\\
29.70$\sim$29.80&145.50&145.11&156.84&-0.40&11.34\\
29.80$\sim$29.90&186.67&183.41&194.42&-3.26&7.75\\
29.90$\sim$30.00&164.27&170.06&185.06&5.79&20.79\\
30.00$\sim$30.10&170.41&180.75&192.99&10.34&22.58\\
30.10$\sim$30.20&190.23&184.10&196.93&-6.13&6.70\\
30.20$\sim$30.30&169.75&178.41&190.35&8.66&20.60\\
30.30$\sim$30.40&162.25&167.36&178.78&5.11&16.53\\
30.40$\sim$30.50&179.54&180.13&189.30&0.59&9.76\\
30.50$\sim$30.60&171.23&172.59&184.03&1.36&12.80\\
30.60$\sim$30.70&150.37&154.99&168.70&4.61&18.33\\
30.70$\sim$30.80&182.95&184.85&195.79&1.90&12.84\\
30.80$\sim$30.90&197.42&202.28&215.48&4.86&18.06\\
30.90$\sim$31.00&158.36&167.72&179.28&9.37&20.93\\
31.00$\sim$31.10&154.39&165.78&177.60&11.39&23.21\\
31.10$\sim$31.20&143.08&146.60&159.93&3.52&16.86\\
31.20$\sim$31.30&159.00&167.74&179.31&8.73&20.31\\
31.30$\sim$31.40&165.01&174.76&187.35&9.76&22.35\\
31.40$\sim$31.50&159.08&168.24&179.00&9.16&19.92\\
31.50$\sim$31.60&168.77&176.52&188.76&7.75&19.99\\
31.60$\sim$31.70&148.84&155.02&169.39&6.18&20.55\\
31.70$\sim$31.80&155.60&161.41&172.84&5.81&17.23\\
31.80$\sim$31.90&147.21&157.40&170.39&10.19&23.17\\
31.90$\sim$32.00&148.55&159.74&171.84&11.19&23.29\\
32.00$\sim$32.10&138.15&149.59&161.64&11.44&23.49\\
32.10$\sim$32.20&164.11&175.00&185.58&10.90&21.47\\
32.20$\sim$32.30&138.54&146.32&154.47&7.78&15.93\\
32.30$\sim$32.40&157.70&168.89&178.56&11.18&20.86\\
32.40$\sim$32.50&163.62&174.65&186.63&11.03&23.01\\
32.50$\sim$32.60&142.35&153.73&165.78&11.38&23.42\\
32.60$\sim$32.70&153.93&165.21&177.38&11.29&23.45\\
32.70$\sim$32.80&141.57&153.35&163.56&11.78&21.99\\
32.80$\sim$32.90&122.22&132.83&142.56&10.61&20.34\\
32.90$\sim$33.00&150.76&161.09&173.66&10.33&22.90\\
33.00$\sim$33.10&147.01&158.06&170.13&11.04&23.12\\
33.10$\sim$33.20&147.19&158.49&170.62&11.30&23.43\\
33.20$\sim$33.30&122.85&133.88&146.34&11.03&23.49\\
33.30$\sim$33.40&148.71&159.70&171.10&10.99&22.39\\
33.40$\sim$33.50&142.98&154.33&166.24&11.34&23.25\\
33.50$\sim$33.60&123.53&135.04&146.56&11.50&23.03\\
33.60$\sim$33.70&137.02&148.75&160.51&11.72&23.48\\
33.70$\sim$33.80&132.94&144.37&156.43&11.42&23.49\\
33.80$\sim$33.90&139.85&150.60&163.34&10.75&23.48\\
33.90$\sim$34.00&167.10&178.13&190.58&11.02&23.48\\
34.00$\sim$34.10&159.45&170.94&182.93&11.49&23.48\\
34.10$\sim$34.20&133.05&144.72&156.21&11.67&23.16\\
34.20$\sim$34.30&155.50&167.17&178.99&11.67&23.48\\
34.30$\sim$34.40&138.34&148.74&160.94&10.40&22.60\\
34.40$\sim$34.50&136.12&147.53&159.46&11.41&23.34\\
34.50$\sim$34.60&165.42&177.18&188.91&11.76&23.49\\
34.60$\sim$34.70&118.65&129.58&142.13&10.93&23.48\\
34.70$\sim$34.80&138.43&147.65&159.51&9.22&21.08\\
34.80$\sim$34.90&162.45&172.96&184.63&10.51&22.19\\
34.90$\sim$35.00&144.51&155.51&167.64&11.00&23.14\\
35.00$\sim$35.10&132.01&142.79&154.73&10.78&22.72\\
35.10$\sim$35.20&155.35&166.68&178.84&11.33&23.48\\
35.20$\sim$35.30&159.64&171.26&183.08&11.62&23.44\\
35.30$\sim$35.40&142.84&154.16&166.32&11.32&23.48\\
35.40$\sim$35.50&124.22&135.79&147.71&11.57&23.49\\
35.50$\sim$35.60&124.13&135.06&146.73&10.94&22.60\\
35.60$\sim$35.70&140.44&150.91&162.92&10.47&22.48\\
35.70$\sim$35.80&130.35&140.85&152.18&10.50&21.84\\
\enddata
\end{deluxetable}

%%%%%%%%%%%%%%%%%%%%%%%%%%%table 3%%%%%%%%%%%%%%%%%%%%%%%%%%%%%%%%%%%%%%%

Assuming $T_{\mathrm{min}}\ll T_{\mathrm{max}}$ and $m<2$, $I(E)$ can be
approximated to
\begin{equation}
I(E)\approx \left\{
\begin{array}{llllllllllllllllllll}
\displaystyle\frac{1}{2-m}\left( \frac{E}{kT_{\min }}\right) , & {E\ll k{%
T_{\min }}} &  &  &  &  &  &  &  &  &  &  &  &  &  &  &  &  &  &  \\
{{\Gamma (3-m)\zeta (3-m)\left( \displaystyle\frac{E}{kT_{\mathrm{min}}}%
\right) ^{m-1}}}, & {k{T_{\min }}\ll E\ll k{T_{\max }}} &  &  &  &  &  &  &
&  &  &  &  &  &  &  &  &  &  &  \\
{\displaystyle\left( \frac{T_{\min }}{T_{\max }}\right) ^{2-m}\left( \frac{E%
}{kT_{\min }}\right) {e^{-\frac{E}{{k{T_{\max }}}}}}}, & {k{T_{\max }}\ll E}
&  &  &  &  &  &  &  &  &  &  &  &  &  &  &  &  &  &  \\
&  &  &  &  &  &  &  &  &  &  &  &  &  &  &  &  &  &  &
\end{array}%
\right.
\end{equation}%
where $\Gamma (x)$ and $\zeta (x)$ are the Gamma function and Zeta function,
respectively. The above equation shows that the low energy ($E\ll kT_{\min }$%
) photon spectrum follows the Rayleigh-Jeans law of $N(E)\propto E$, while
the high energy ($E\gg kT_{\max }$) spectrum behaves as a power law with an
exponential cutoff. For the middle regime of the spectrum ($kT_{\min
}<E<kT_{\max }$), different $m$ corresponds to different shape. We can see
that $m\leq 2$ is required. Otherwise, the photon spectrum $N(E)\propto
E^{m-1}$ in the energy band of $kT_{\min }<E<kT_{\max }$ is harder than the
blackbody $N(E)\propto E$, which is physically unrealistic. In fact, for $%
m>2 $, $N(E)$ is always proportional to $E$ for $E<kT_{\max }$, as in this
case the whole spectrum is dominated by the $T=T_{\max }$ blackbody. Thus
compared with Ryde et al. (2010), by considering the effect of $T_{\min }$,
we obtain the complete mBB spectrum including the Rayleigh-Jeans
component ($E<kT_{\min }$).

In Figure 1, we plot the mBB spectra for different $m$
values. $kT_{\min}=1$ keV and $kT_{\max}=100$ keV are assumed. For $%
m=2$, the spectrum is similar to a pure blackbody. In the $E^{2}N(E)$
spectrum, however, it deviates from the pure blackbody by a factor of $\ln
(kT_{\max }/E)$ from $kT_{\min }$ to $kT_{\max }$. Note that for $
m=2$ the low energy spectrum of $E<3kT_{\min}$ is influenced by the entire black body components with temperatures from $kT_{\min}$ to $kT_{\max}$, while the low energy spectrum is mainly dominated by the black body component with temperature $\sim kT_{\min}$ for $m<2$, which explains a higher $E^2 N(E)$ for the $m=2$ case. The spectrum in the case
of $m=1$ can nicely explain the observed narrowly-shaped spectrum of GRB 090902B. For $m=0$, it is the same as the Comptonized spectrum, which is
also known as the COMPTEL model (Goldstein et al. 2012). As for GRB 081221
analyzed in the next section, the spectrum ($m=-0.46$) is between the two
cases of $m=0$ and $m=-1$ . If the minimum and maximum temperatures are
close enough, i.e., $T_{\min }\sim T_{\max }$, the spectrum returns to the
pure blackbody. On the other hand, if the minimum temperature $T_{\min }$ is
below the energy band of the GRB detector, the low energy part of the mBB
spectrum can not be observed, and the middle and the high energy parts might
be approximated as the well-known Band spectrum or the Comptonized cut-off
power law spectrum in a limited energy band determined by the detector. In this paper, we only consider the case of a single
power-law distribution of the blackbody temperature. The temperature
distribution of more realistic GRB fireballs may be much complicated.

\section{Spectral Analysis of GRB 081221\label{sec:data}}

GRB 081221 at redshift $z=2.26$ (Salvaterra et al. 2012), a bright GRB with
separable pulses, was \textbf{detected} by Fermi/GBM, Swift/BAT, and
Konus/Wind. There was no detection in the LAT band. We analyze the GBM data
using the same method as described in Zhang et al. (2011) and fit the
extracted spectra with McSpecFit package (Zhang et al 2016). We selected the
data from GBM detectors NaI 1, NaI 2 and BGO 0, according to their geometry
relative to the GRB position. The light curve with a duration of about 40 s
consists of a weak initial peak (precursor) and a bright second peak (main
burst), as can be seen in Figure  2. The duration of the precursor is about 5
s, and the quasi-quiescent time is $\sim $ 10 s between these two peaks. The
main burst consists of three overlapping pulses with a duration of $%
\sim $ 20 seconds (Cummings et al. 2008; Golenetskii et al. 2008;
Wilson-Hodge 2008). The event fluence of GRB 081221 is $(3.7\pm
0.1)\times 10^{-5}\mathrm{~erg~cm^{-2}}$ (8-1000 keV), and the isotropic energy is $%
(4.21\pm 0.42)\times 10^{52}\mathrm{~erg}$.

%%%%%%%%%%%%%%%%%%%%%%%%%%Figure 2%%%%%%%%%%%%%%%%%%%%%%%%%%%%%%%%%%%%%%%

\begin{figure}[tbp]
\label{Fig_2} \centering%
\includegraphics[angle=0,scale=0.2]{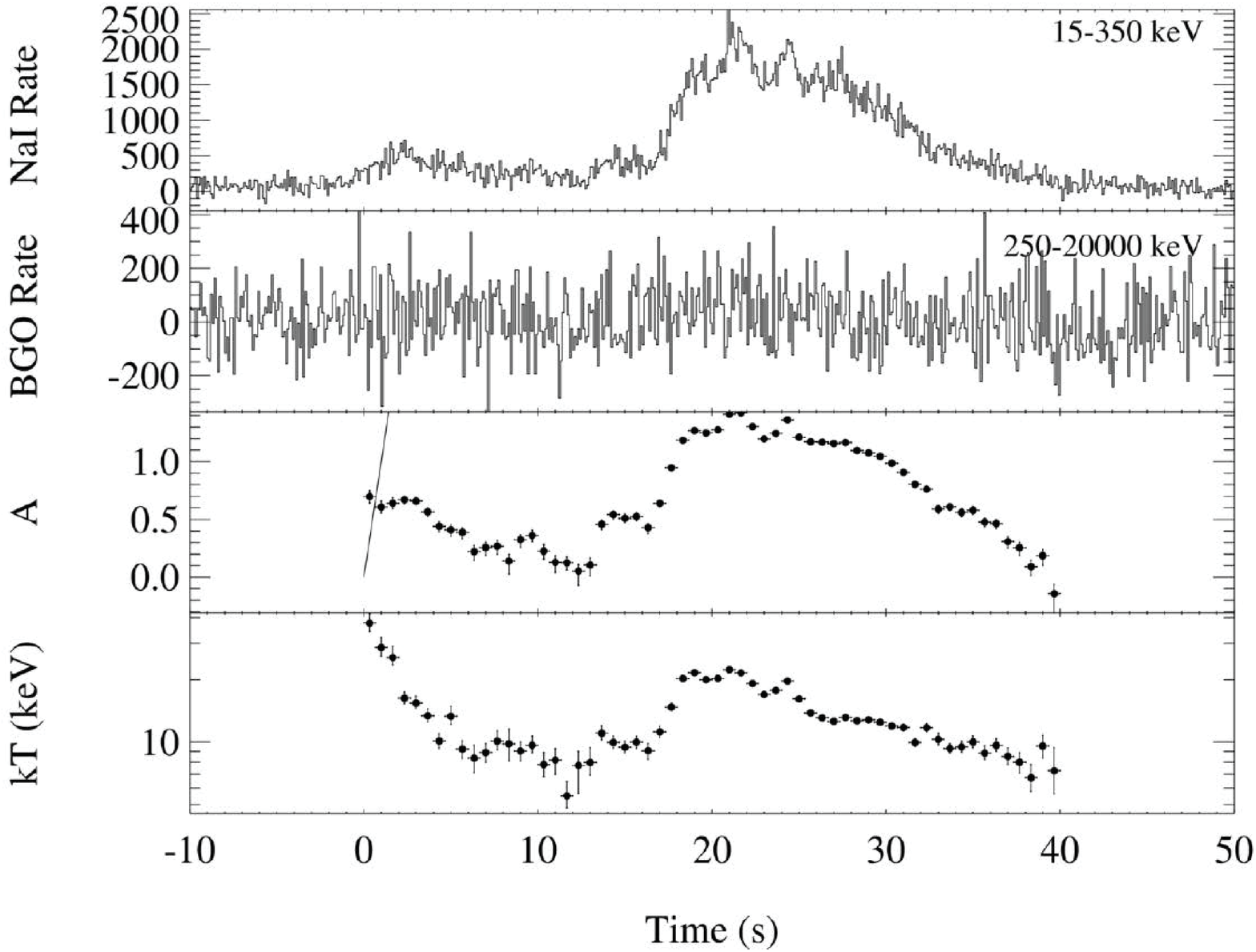}
\caption{GBM light curves of GRB 081221 (top two panels), and the
spectral evolution (bottom two panels). }
\end{figure}

%%%%%%%%%%%%%%%%%%%%%%%%%%%%%%%%%%%%%%%%%%%%%%%%%%%%%%%%%%%%%%%%%%%%

We perform the time-resolved spectral analysis of GRB 081221 using the GBM
data. We use the personal IDL code library ZBBIDL and personal Python library ZBBPY in
the whole data processing project developed and applied in Zhang et al. (2018).
Divided the whole burst into 40 slices at first, we use a single blackbody (BB) model
and an mBB model to fit the data, respectively. The fitting results are presented in Table 1.
The Bayesian Information Criterion (BIC) is a tool for model selection (Schwarz 1978; Wei et al. 2016).
A model is preferred when it has the lower BIC value than the other. We define $\Delta \rm{BIC}$ = $\rm{BIC}_{1}$ - $\rm{BIC}_{2}$, where $\rm{BIC}_{i}$ corresponds to Model $i$ ($i=1, 2$).
If $\Delta \rm{BIC}$ is from 2 to 6, the preference for Model 2 is positive; if $\Delta \rm{BIC}$ is from 6 to 10, the preference for Model 2 is strong;
and if $\Delta \rm{BIC}$ is above 10, the preference for Model 2 is very strong. By comparing their BIC,
we come to our conclusion that although the BB model is not the best-fit model,
we can use this model to explore the physical properties of the GRB photosphere, and the BB model
can be reasonably used to fit the data (Zhang et al. 2018). We also find when the weaker the radiation is,
the more suitable the single blackbody model is. This means that the time slice with high fluence may be superimposed
by a number of the BB components. The spectral evolution may be another important reason.
We are then motivated to take a smaller time bin.
Thanks to the high flux of this burst, we divide the burst into 250 slices based on the criterion that there are at least 20 net counts in at least 10 energy channels in the spectrum to conduct a fine-time-resolved spectral analysis.
The two blackbodies plus a power-law model (2BBPL), which was applied to GRB 081221 by Basak \& Rao (2013), is also considered in this work.
The computed BIC for the time-resolved spectra of GRB 081221 with these three models are listed in Table 3.
The evolution diagram of the $\Delta \rm{BIC}$ with time and histogram about $\Delta\rm{BIC}_{mBB - BB}$ and $\Delta\rm{BIC}_{2BBPL - BB}$
are shown in Figure 3. The $\Delta \rm{BIC}$ of most time intervals is larger than 10, so it can be judged that
the single blackbody model is the best-fit model.
Comparing the BIC values of the BB model in the  Tables 1 and 3,
we also find that the smaller the interval is, the more acceptable the single blackbody model is.
It means that the whole radiation is superimposed with blackbodies of different temperature, and the spectral evolution has a great influence on the time-integrated spectrum.
The temporal evolution of ${kT}$ is shown in
Figure 2. The $kT$ of the precursor evolves quickly from $\sim 30$ keV to $%
\sim $ 7 keV. In the main part of the burst, $kT$ increases rapidly from $\sim$ 9
keV to $\sim$ 20 keV, then decreases with the flux.

%%%%%%%%%%%%%%%%%%%%%%%%%%Figure 3%%%%%%%%%%%%%%%%%%%%%%%%%%%%%%%%%%%%%%%

\begin{figure}[tbp]
\includegraphics[angle=0,scale=0.8]{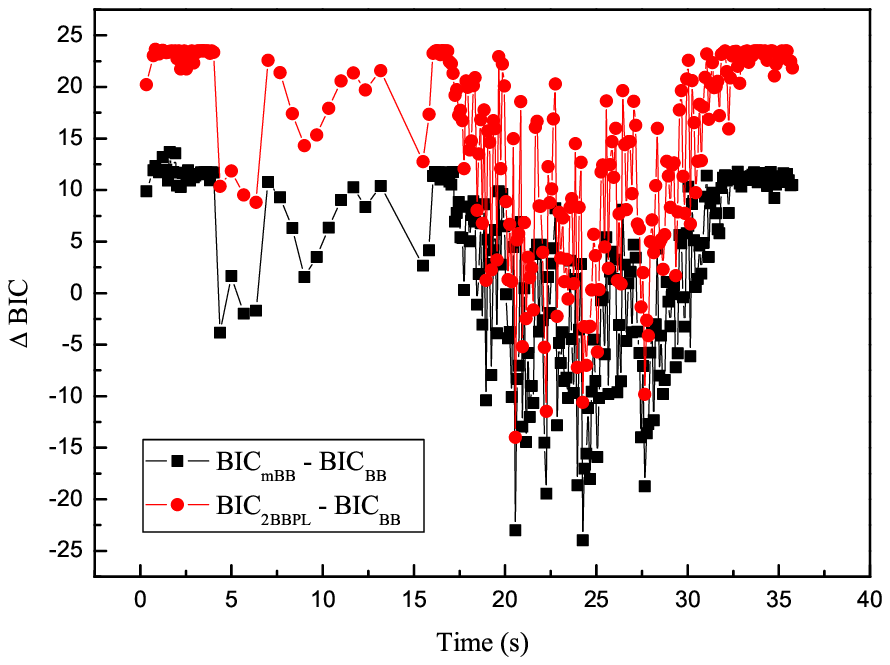}
\includegraphics[angle=0,scale=0.8]{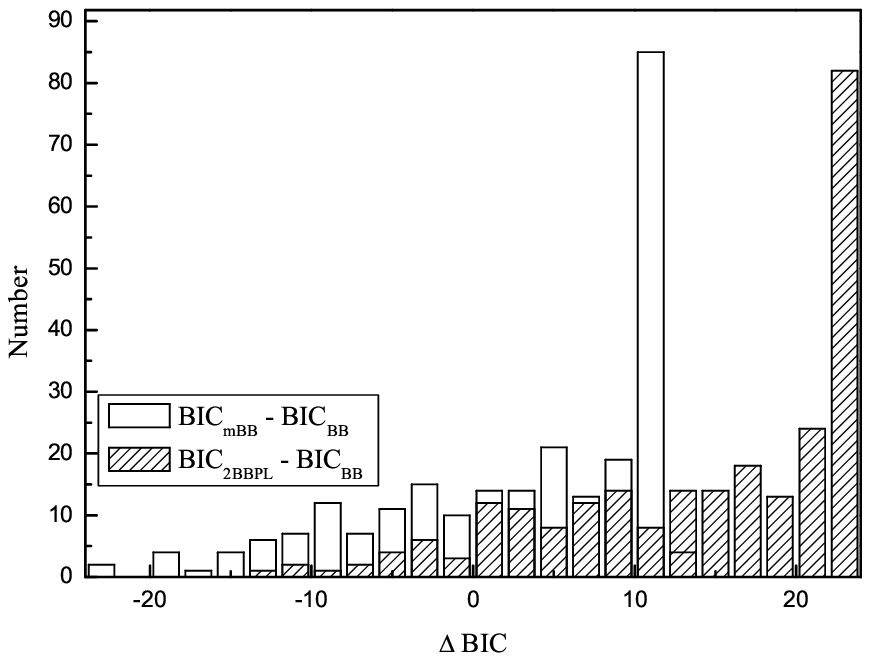}
\caption{The evolution of the $\Delta$BIC with time for the three models (left panel), and histogram of the $\Delta$BIC (right panel ). }
\end{figure}

%%%%%%%%%%%%%%%%%%%%%%%%%%%%%%%%%%%%%%%%%%%%%%%%%%%%%%%%%%%%%%%%%%%%

The left panel of Figure 4 shows an example of a spectral fitting with a BB
model in the interval of [19.5s, 20.0s], when the prompt emission is around
the brightest peak, where we obtained $kT=19.9\pm 0.5$ keV. We noted the
non-Gaussian pattern of the residuals indicate that the spectral evolution
may still exist even within such a small time interval. Furthermore, the
right panel of Figure 4 shows the flux - $kT$ correlation for the
time-resolved spectra, a higher flux generally corresponds to a BB component
with a higher temperature (Golenetskii et al. 1983; Kargatis et al. 1994;
Borgonovo \& Ryde 2001; Fan et al. 2012). This is similar to the ${E_{p}}$%
-flux correlation in the Band model fittings (Lu et al. 2010). We
fit the flux$-kT$ data with a power-law for the decay phase of the second pulse ($t>$ 30s) and the time interval of 12 - 40s.
The slopes are $\sim 4.1$ and $\sim 2.7$, respectively.
For the decay phase of the second pulse, this slope is consistent
with the BB model prediction. For a BB, the luminosity
of the photospheric emission is $L=16\pi R^{2}\frac{(1+z)^{4}}{\Gamma ^{2}}%
\sigma T^{4}$, where $R$ is the photospheric radius and $\Gamma $ is the
Lorentz factor. The luminosity is less sensitive to the radius ($\propto R^2$%
) and the Lorentz factor ($\propto\Gamma^{-2}$) than to the temperature ($%
\propto T^4$), so it is expected that the flux approximately proportional to
$T^4$. The deviation of the observed dependence of the BB flux on
temperature ($T^{2.7}$) during the time interval of 12 - 40 s to the theoretical $T^{4}$ should be
attributed to some mild temporal evolution of the photosphere radius and/or
the Lorentz factor, as this time interval corresponds to the whole period of the main burst of GRB 081221, which consists of the rising, peak and decaying parts. Note that the slopes of 4.1 and 2.7 here are irrelevant
with the power-law index $m$ in Eq. (3).

The model fittings to the time-integrated GBM spectrum between 20 and 30 s
of GRB 081221 are shown in Figure 5. Two spectral breaks at $\sim 15$ keV
and $\sim 180$ keV are seen in this figure. The spectrum rises fast below $%
\sim 15$ keV, then shows a plateau segment between $\sim 15$ keV and $%
\sim 180$ keV, and finally follows a steep decrease above $\sim 180$ keV.
The fitting results are presented in Table 2. Parameter constraints of the
mBB model and Band function model for time-integrated spectrum between 20
and 30 s are shown in Fig.6. We find that the multi-color blackbody model ($%
m=-0.46$) can fit almost as well as the Band model, which means that the
Band spectrum in GRB 081221 can be well interpreted as the mBB.
The derived minimum and maximum temperatures ( $kT_{\min }=4.4\pm
0.3$ keV, $kT_{\max }=57.0_{-1.4}^{+1.6}$ keV) for the time-integrated
spectrum. The derived single blackbody temperatures range from $%
12.45^{+0.53}_{-0.46}$ keV to $22.38^{+0.51}_{-0.46}$ keV. This difference
can be understood by the following factors: (1) Different formulation. $%
T_{\min}$ and $T_{\max}$ can be regarded as effective temperatures which are
not exactly the term of the temperature $kT$ in the BB model.
(2) Different selection of data. The first approach is applied to the whole
time-integrated spectra while the 2nd approach is applied to each slice of
the time-resolved spectra. So $T_{\min}$ and $T_{\max}$ in mBB are
``averaged" values and are subject to spectral evolution. Thus they are not
necessarily exactly as same as $kT$ in the BB model.

\ \ Based on the above spectral analysis, we find that the prompt emission
of GRB 081221 may be overwhelmed by the photospheric (thermal) emission,
without any significant non-thermal contribution in the GBM energy band.

\section{Discussion and Conclusions}

Usually, the spectrum of thermal emission from astronomical objects can be
described by the Planck function. However, for GRBs which are widely
believed to have a relativistic outflow, due to the angle dependence of the
photospheric emission together with the Doppler boosting effect (Pe'er
2008), the photospheric emission should be a superposition of blackbodies of
different temperatures, which is also called the mBB.

We notice that, Ryde et al. (2010) and Larsson et al. (2011) used a
broadened photospheric component based on an mBB model and
the XSPEC model diskpbb to fit the spectra of GRB 090902B and GRB 061007,
respectively. In their works, they adopted $T_{\min}\simeq 0$ as $T_{\min }$
$\ll $ $T_{\max }$ and $kT_{\min}$ below the lowest energy of the energy band of the detector in their cases. In the diskbb model, the observed photon
flux is given by
\begin{equation}
N(E)=\int_{{T_{\mathrm{out}}}}^{{T_{\mathrm{in}}}}{(\frac{T}{{{T_{\mathrm{in}%
}}}}}{)^{m}}B(E,T)\frac{{dT}}{{{T_{\mathrm{in}}}}},
\end{equation}
where the temperature of the outer edge of the disk can be regarded as ${T_{%
\mathrm{out}}}=0$. So the diskbb model is effectively the same as the model
of Ryde et al. (2010), in which the temperature distribution is normalized
by $T_{\mathrm{max}}$ ($T_{\mathrm{in}}$ in the diskbb model).

It is worth pointing out that our model is slightly different from Ryde et
al. (2010). In Ryde et al. (2010), they assumed a \textbf{probability distribution}
of the mBB temperature. In our model, the distribution of
the photosphere emission luminosity as a function of temperature is more
directly related to the physics of the GRB fireball, and the power law index
$m$ of the distribution normalized by $T_{\mathrm{min}}$ decides the shape
of the low and middle energy part of the spectrum. If $T_{\mathrm{min}}$ is
small enough as to be outside the low-energy limit of the instrument,
the observed spectrum becomes the Band or Comptonized cutoff power-law
spectrum. Otherwise, if $T_{\mathrm{min}}$ is high enough as to be
comparable to the $T_{\mathrm{max}}$, the observed spectrum is close to a
pure blackbody.

In this paper, we systematically analyze the time-integrated and
time-resolved spectra of the prompt emission of GRB 081221 based on the
Fermi/GBM data. We find that the time-integrated
spectrum is well fitted by the mBB model and the
time-resolved spectra can be well fitted by the BB model
(without any non-thermal power-law component), implying that the prompt
emission of GRB 081221 may be dominated by the photospheric emission. Our
results are summarized as follows:

\begin{itemize}
\item The time-integrated spectrum during the bright phase (from 20s to 30s since the trigger) of GRB 081221 observed with Fermi/GBM has a broad
plateau in the 15-180 keV band, and is well fitted by the mBB
 model, yielding the $kT_{\min }=4.4\pm 0.3$ keV, $kT_{\max
}=57.0_{-1.4}^{+1.6}$ keV, ${m=-0.46_{-0.06}^{+0.05}}$.

\item As for the time-resolved spectra, although the BB model is not
the best-fit model when the whole burst is divided into 40 slices, we obtain acceptable
fits with the single blackbody model (without the power-law component), which is simpler
compared to the 2BB plus PL model of Basak \& Rao (2013). The $kT$ of the
main burst increases rapidly from $\sim$ 9 keV to $\sim$ 20 keV.
When the burst is divided into 250 slices, the BB model is much preferred than the Band model and the 2BB plus PL model, based on the Bayesian Information Criterion.
This means that the evolution of the spectrum has a great influence on the time-integrated spectrum,
and hence the time-binning greatly affects the  theoretical interpretation of the observed time-integrated and even time-resolved spectra.
\end{itemize}

Our results indicate that the mBB should be the main
intrinsic spectral component of GRB 081221, which is a superposition of a
series of blackbodies with different temperatures arising at different times
of the prompt phase. The identification of the photosphere origin of GRB
081221 suggests that the jet composition of this burst is a matter dominated
fireball. On the other hand, the spectra of some bursts such as GRB 130606B
(Zhang et al. 2016) are adequately modeled by the fast cooling synchrotron
radiation model (Uhm \& Zhang 2014), with no evidence of a thermal
photosphere emission component, which suggests a Poynting-flux-dominated jet
composition. The co-existence of both types of GRBs suggests that GRB jet
composition is likely diverse, with different bursts having different
degrees of magnetization from the central engine (e.g. Gao \& Zhang 2015).

\acknowledgments

We appreciate the referee very much for his/her valuable comments and
suggestions. We acknowledge the use of the public data from the Fermi data
archives. This work was supported by the National Basic Research Program
(973 Program) of China (Grants 2014CB845800), the National Natural Science
Foundation of China (Grant Nos 11725314, 11673068, 11473022, 11503011,
11275057, 11543004, 11233006 and 11603006), the Open Research Program of Key
Laboratory for the Structure and Evolution of Celestial Objects (OP201403 and OP201703),
key scientific and technological project of Henan province (No.172102310334),
the Key Research Program of Frontier Sciences (QYZDB-SSW-SYS005) and the
Strategic Priority Research Program ``multi-waveband gravitational wave
Universe'' (grant No. XDB23000000) of the Chinese Academy of Sciences, the
Fundamental Research Funds for the Central Universities, Nanyang Normal
University Doctoral Fund(15064) and the Key Scientific Research Project in
Universities of Henan Province (Grant 15A160001). Part of this work used
BBZ's personal IDL code library ZBBIDL and personal Python library ZBBPY.
The computation resources used in this work are owned by Scientist Support
LLC.

\begin{figure*}[tbp]
\label{Fig_3}
\begin{tabular}{cc}
\includegraphics[angle=0,width=8cm,height=7.5cm]{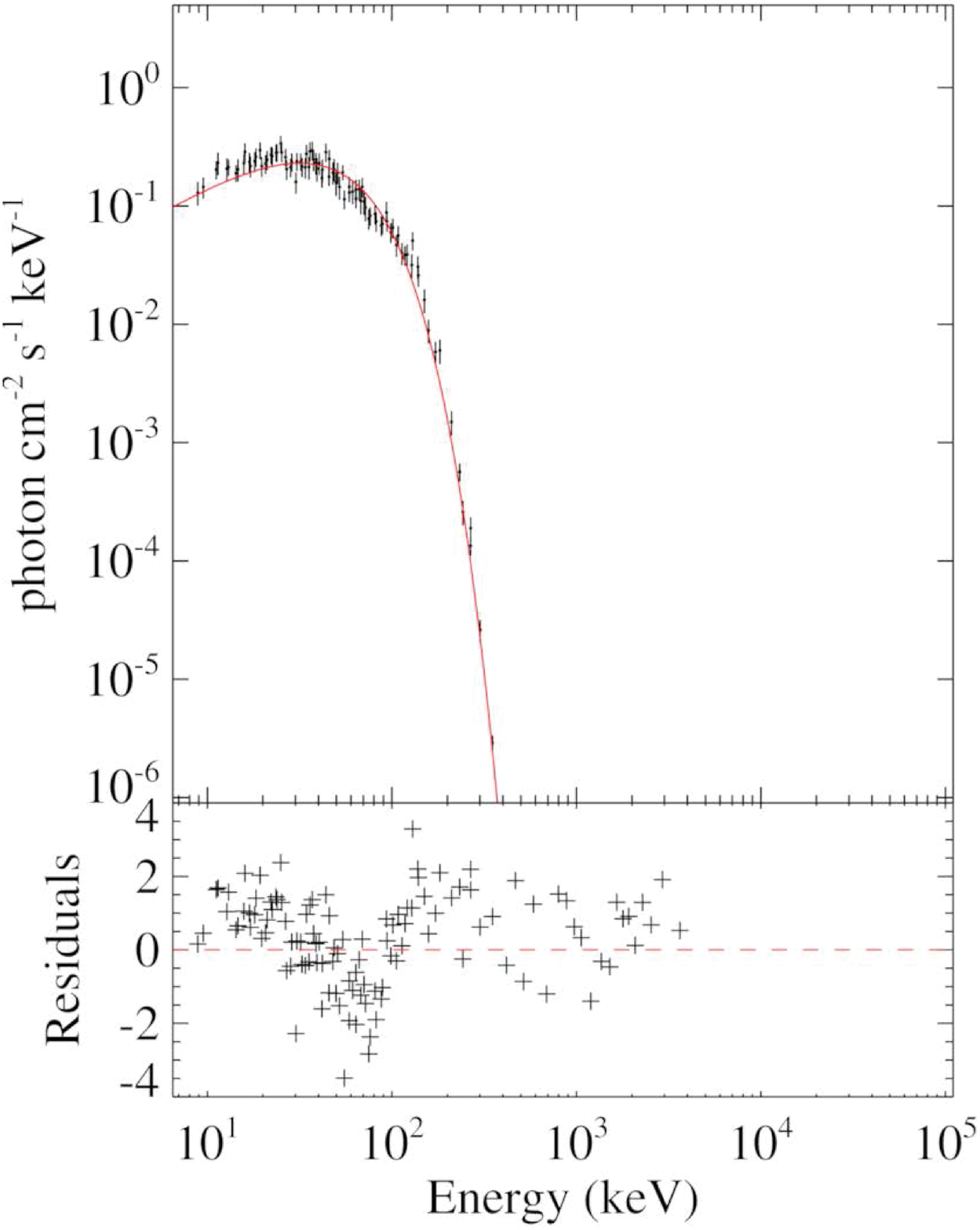} & %
\includegraphics[angle=0,width=8cm,height=7cm]{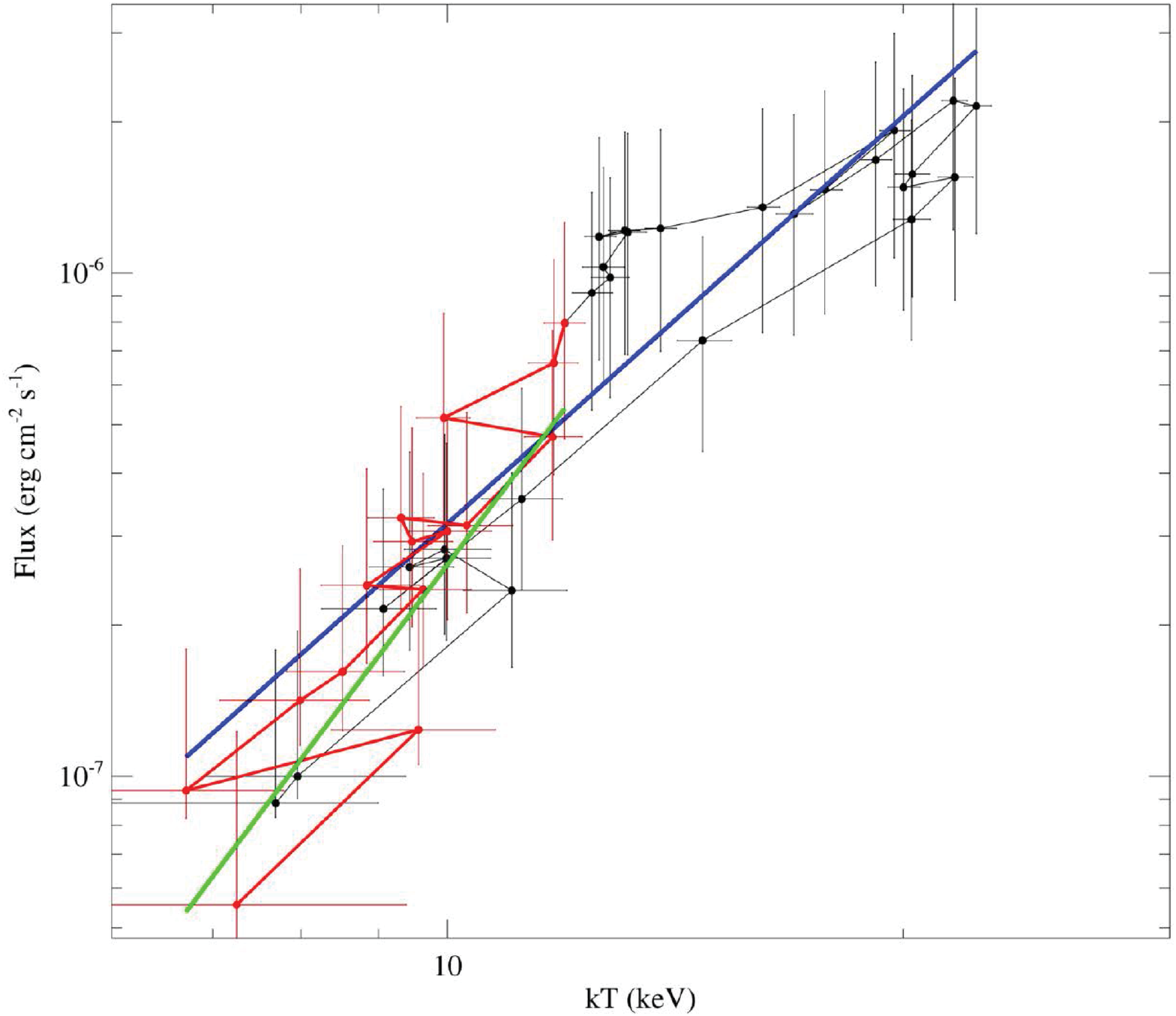} \\
&
\end{tabular}%
\caption{Left: the time resolved photon spectrum when the flux is around the
highest peak (time from 19.5 s to 20.0 s since the trigger), in which a
blackbody model is applied. Right: evolution of $kT$ vs flux. The red points
and green solid line indicate the data and the flux $\propto (kT)^{4.1}$ correlation in the
decay phase of the second (main) pulse ($t>$ 30 s). The blue solid line indicates the flux $\propto (kT)^{2.7}$ for the whole period of the second (main) pulse ($t=12 - 40$) s.}
\end{figure*}

\begin{figure*}[tbp]
\centering
\centering
\begin{tabular}{cc}
\includegraphics[scale=0.2]{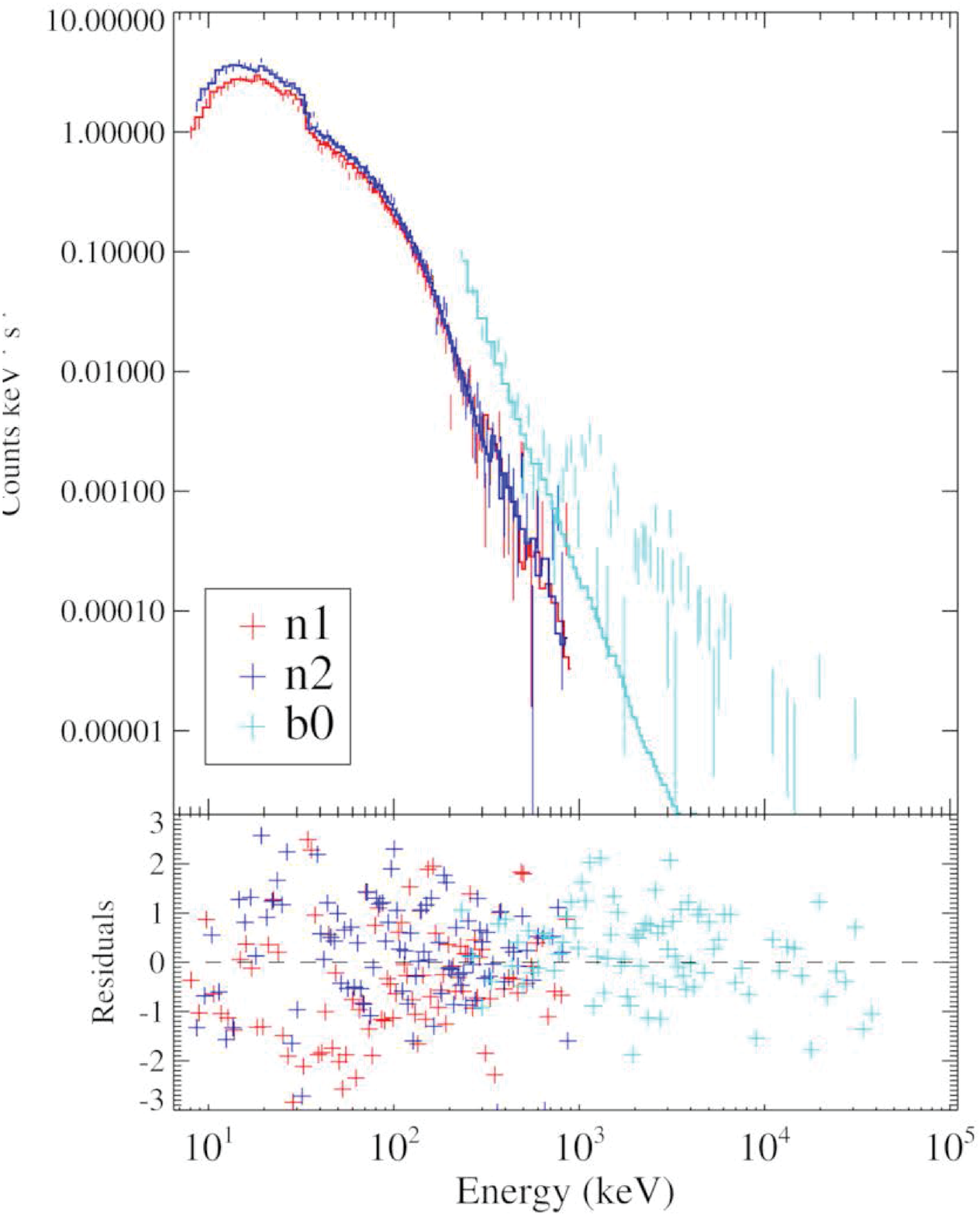} & %
\includegraphics[scale=0.2]{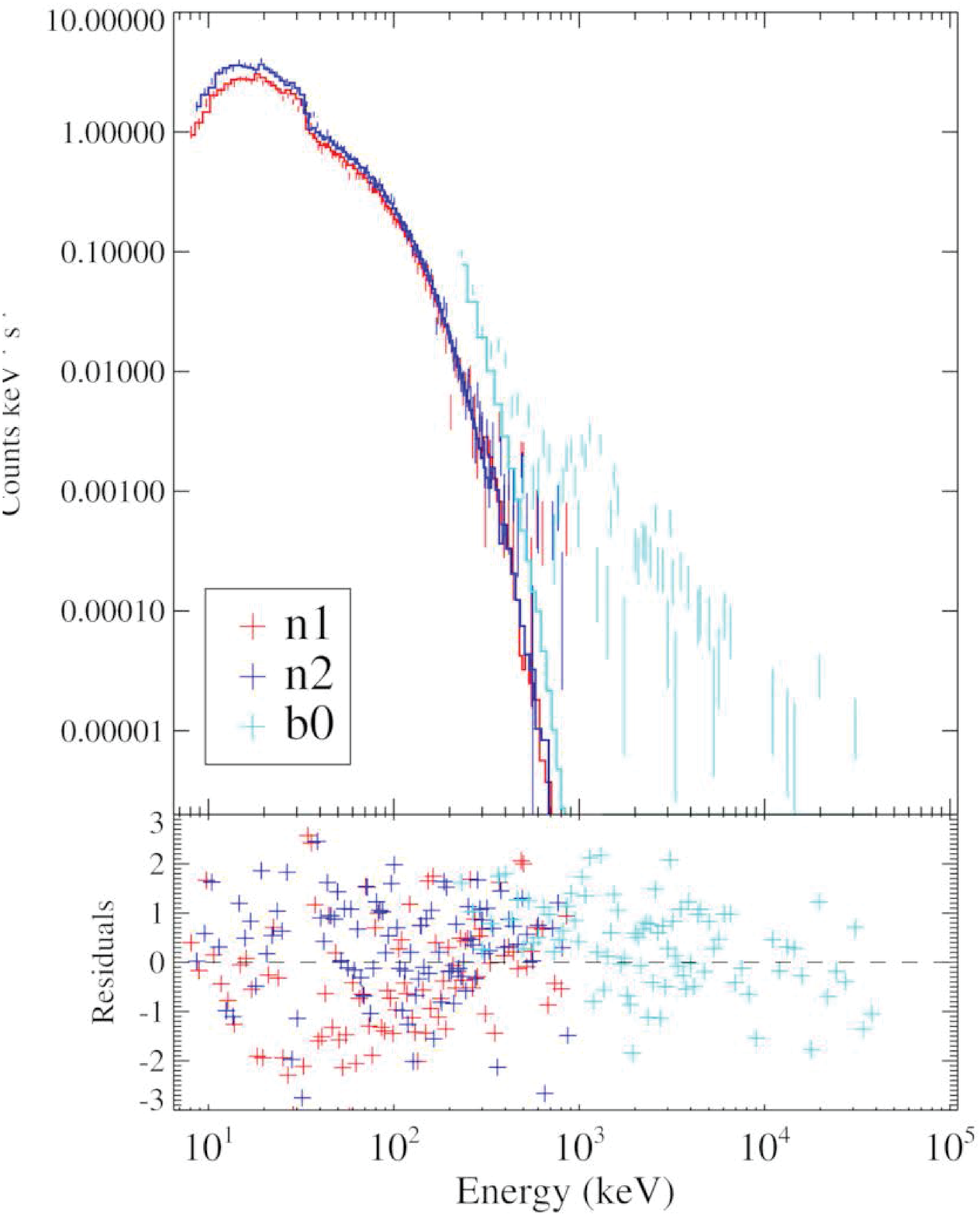} \\
\includegraphics[scale=0.2]{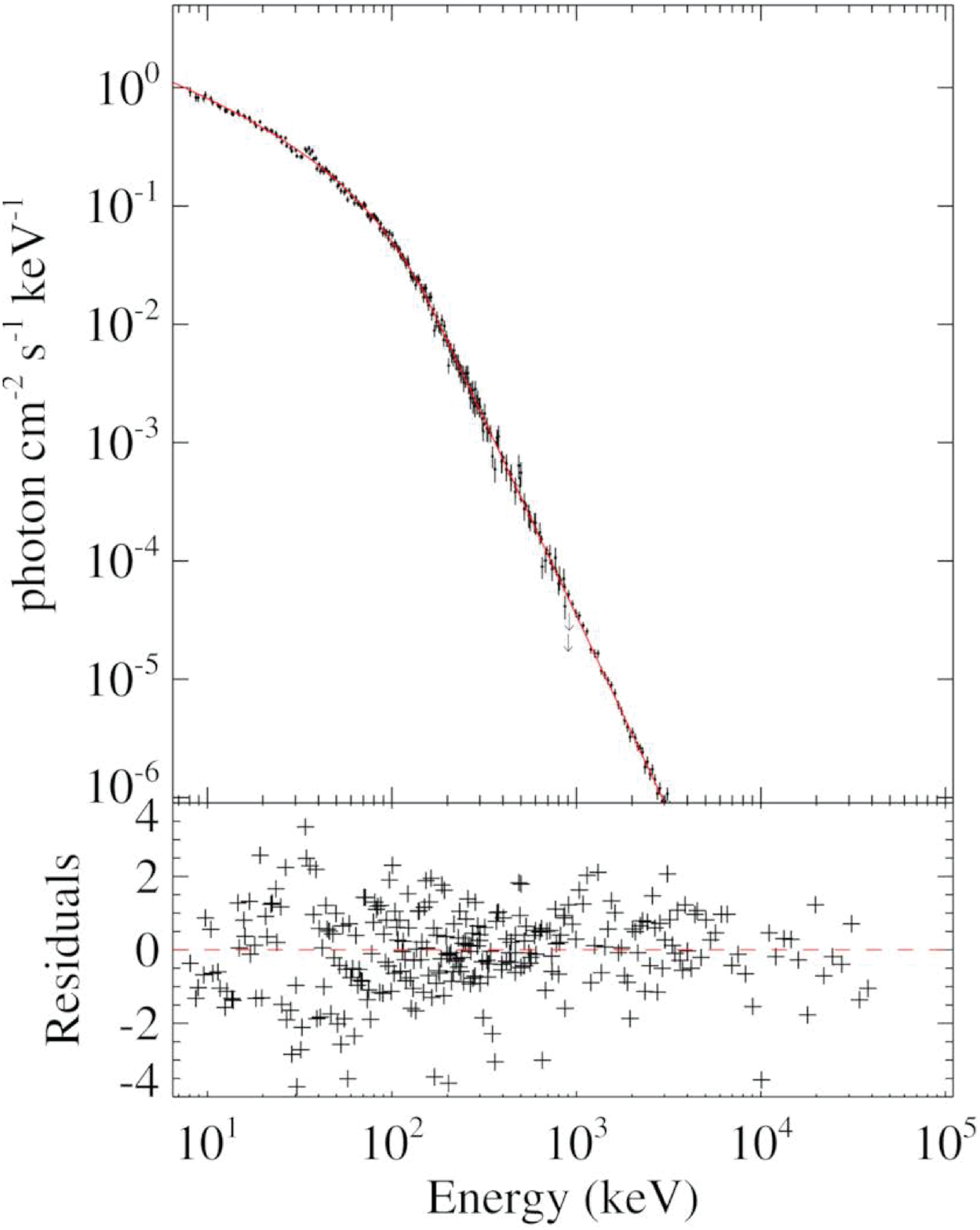} & %
\includegraphics[scale=0.2]{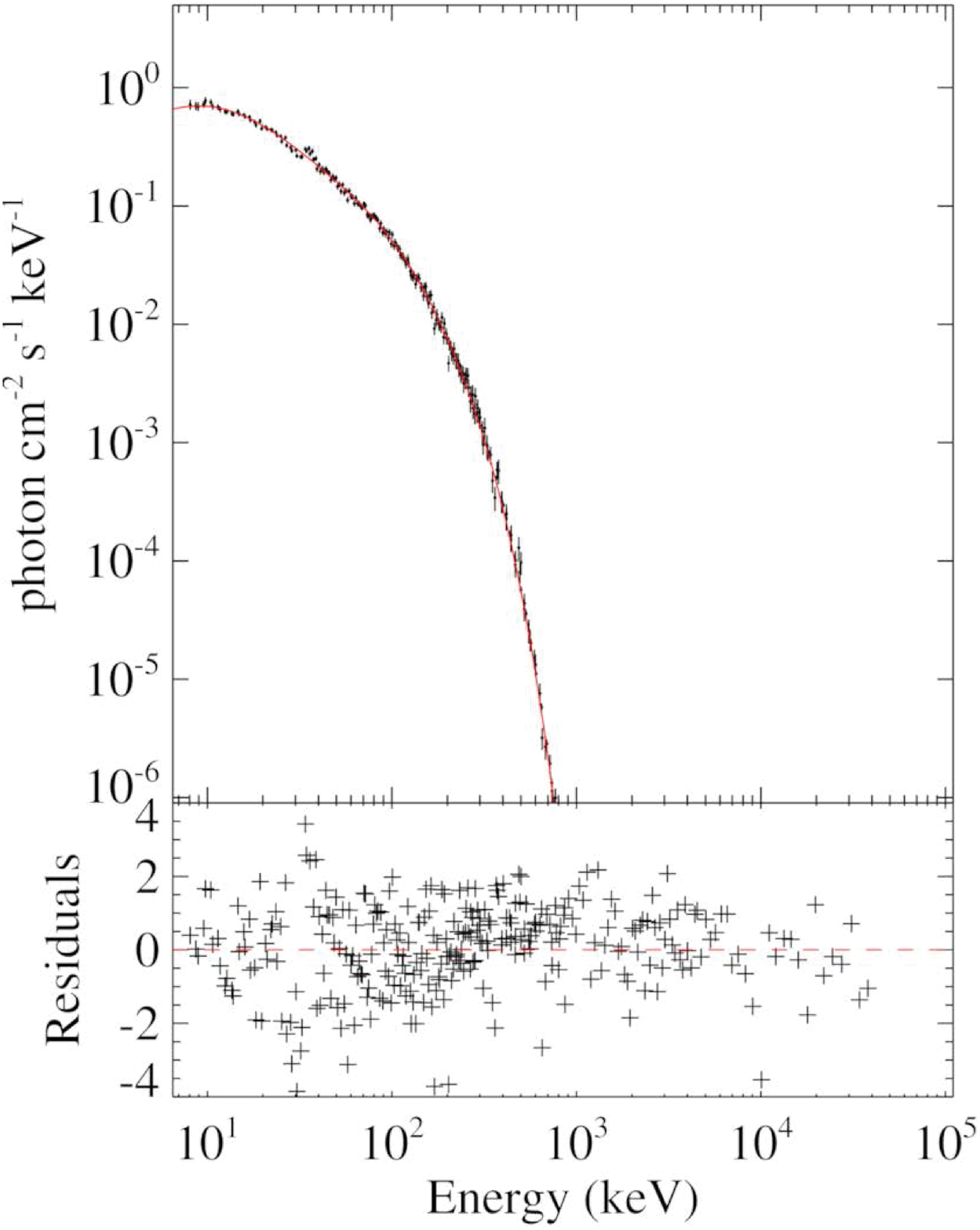} \\
&
\end{tabular}%
\caption{Comparison between the Band function fitting and the mBB model
fitting for the time integrated spectrum between 20 and 30 s. The spectra
have been re-binned into wider bins for clarity. \textit{Top two:} observed
count spectrum vs. model. \textit{Top Left} shows the Band function fitting
and \textit{Top Right} shows the mBB model fitting. \textit{Bottom two:} the
de-convolved photon spectrum plots. \textit{Bottom Left} shows the Band
function fitting, red line is the theoretical photon spectrum of the Band
function and data points are the ``observed" photon flux which is obtained
by de-convolving the observed count spectrum using instrument responses.
\textit{Bottom Right:} same as \textit{Bottom Left} but for the mBB model
fitting where the red line is the theoretical photon spectrum of multi-color
blackbody (mBB) model.}
\label{fig:ph_obs_comp}
\end{figure*}

\begin{figure*}[tbp]
\centering
\begin{tabular}{cc}
\includegraphics[width=3.5in]{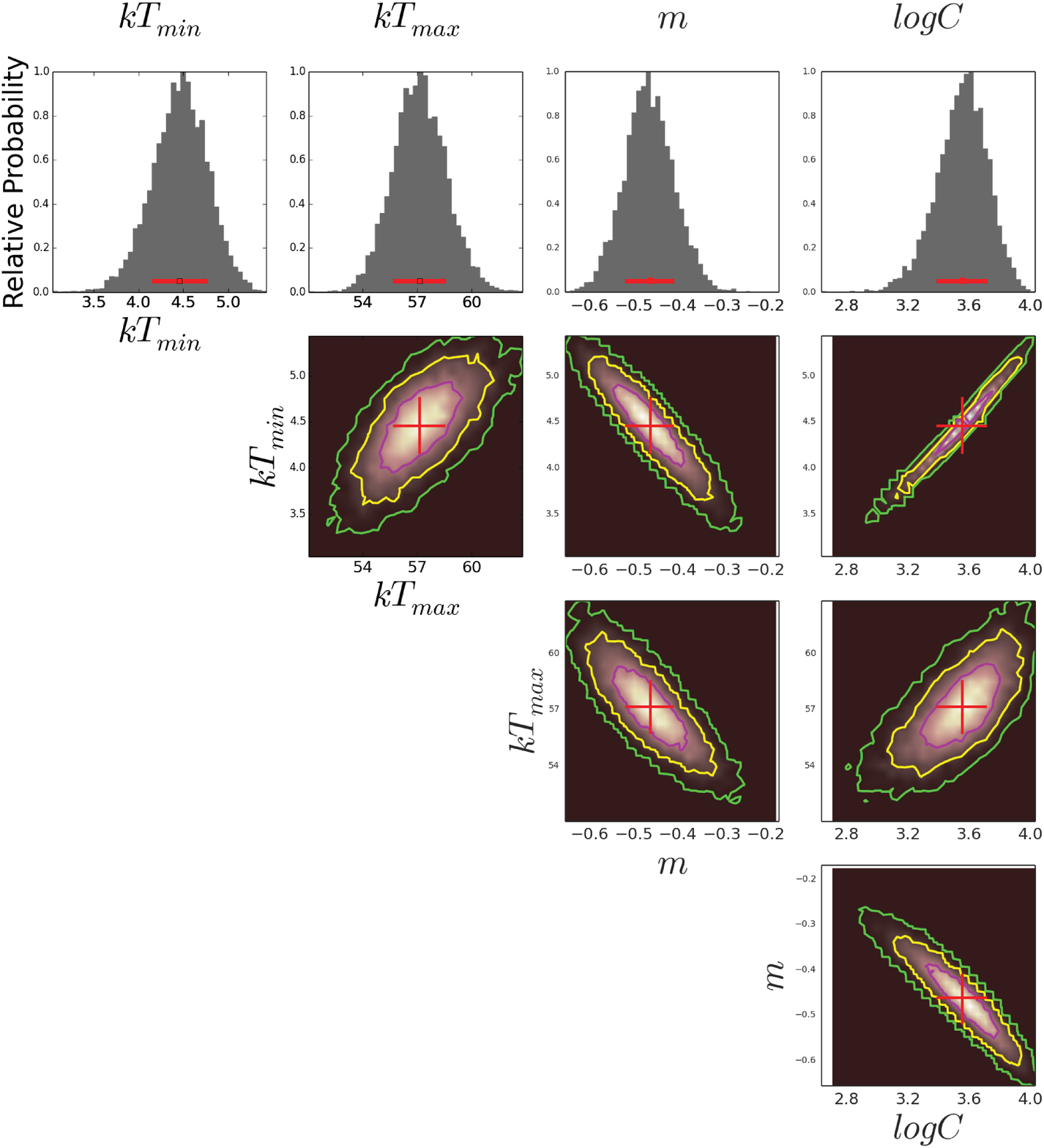} & %
\includegraphics[width=3.5in]{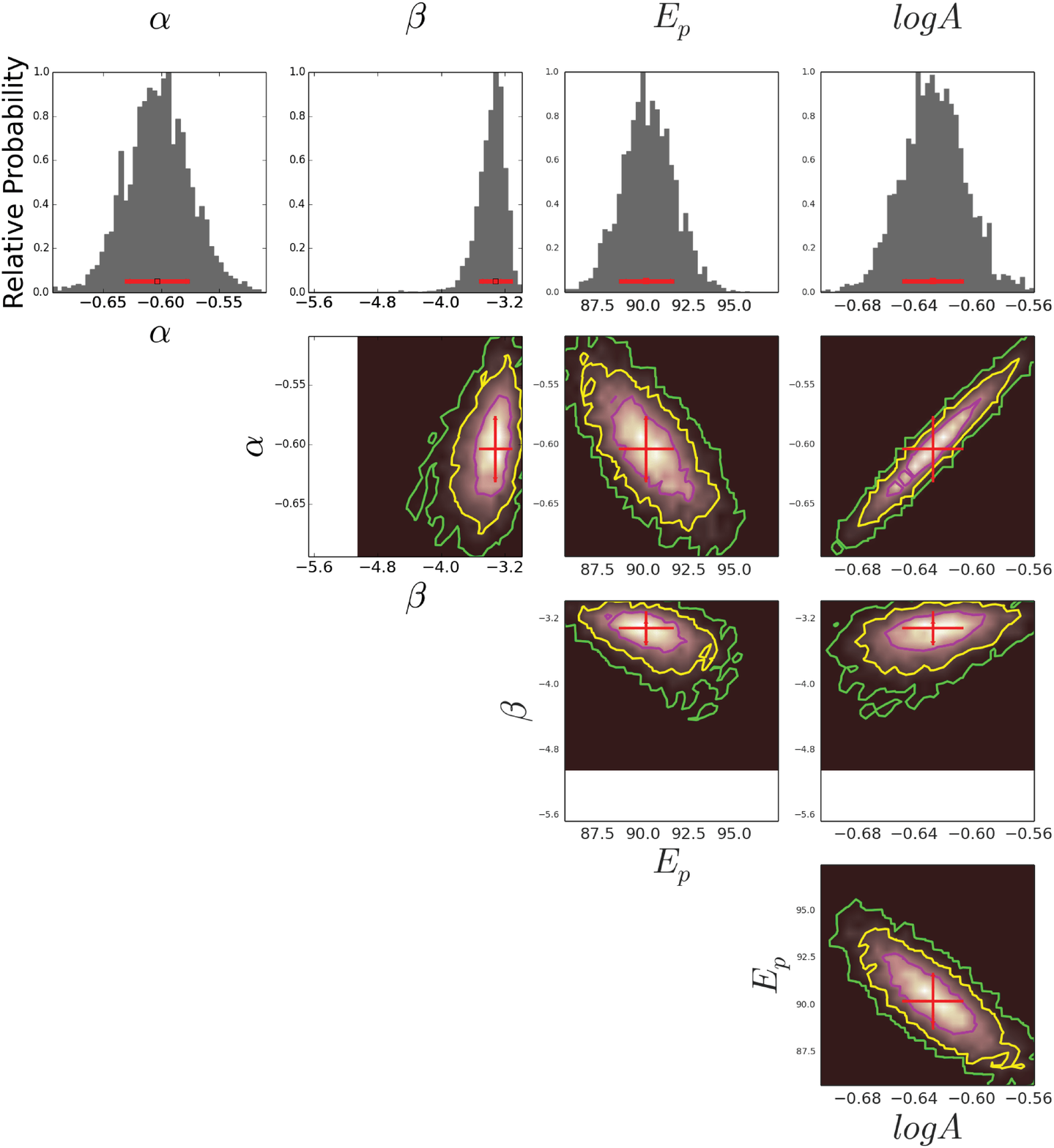} \\
&
\end{tabular}%
\caption{Parameter constraints of the mBB model (\textit{left}) and Band
function model (\textit{right}) for time integrated spectrum between 20 and
30 s. Histograms and contours show the likelihood map of the
parameter-constraint outputs by our \textit{McSpecFit} package. Red crosses
mark the best-fit values and 1-sigma error bars. }
\label{fig:contour}
\end{figure*}

\end{document}